\documentclass[fleqn,useAMS,usenatbib]{./mn2e}
\usepackage{graphicx}
\usepackage{lscape}
\usepackage{times}
\usepackage{amsmath}
\usepackage{amsfonts}
\usepackage{amssymb}
\usepackage{multirow}
\usepackage{longtable}
\usepackage{afterpage}
\usepackage{pdflscape}
\RequirePackage{color}

\title[Fundamental properties of targets]{
Fundamental properties of $Kepler$ and $CoRoT$ targets: IV. Masses and radii from frequencies of minimum  $\Delta \nu$
and their implications
}
\author[M. Y\i ld\i z, Z. \c{C}elik and C. Kayhan]{M. Y\i ld\i z\thanks{E-mail:
mutlu.yildiz@ege.edu.tr}, Z. \c{C}elik Orhan and C. Kayhan \\
Department of Astronomy and Space Sciences, Science Faculty, Ege University, 35100, Bornova, \.Izmir, Turkey.}

\date{Accepted 2019 August 5. Received 2019 July 23; in original form 2019 February 20}

\begin{document}

\pagerange{\pageref{firstpage}--\pageref{lastpage}} \pubyear{2013}
\def\braket#1{\left<#1\right>}
\newcommand{\yildiz}{Y\i ld\i z }
\newcommand{\etal}{et al. }
\newcommand{\wrt}{with respect to }
\newcommand{\logg}{\log(g) }
\newcommand{\numino}{\mbox{\ifmmode{\overline{\nu_{\rm min}}}\else$\overline{\nu_{\rm min}}$\fi}}
\newcommand{\numin}{\mbox{\ifmmode{\nu_{\rm min}}\else$\nu_{\rm min}$\fi}}
\newcommand{\teff}{\mbox{\ifmmode{T_{\rm eff}}\else$T_{\rm eff}$\fi}}
\newcommand{\teffsun}{\mbox{\ifmmode{{\rm T}_{\rm eff\sun}}\else${\rm T}_{\rm eff\sun}$\fi}}
\newcommand{\numax}{\mbox{$\nu_{\rm max}$}}
\newcommand{\nuH}{\mbox{\ifmmode{\nu_{\rm minH}}\else$\nu_{\rm minH}$\fi}}
\newcommand{\nuL}{\mbox{\ifmmode{\nu_{\rm minL}}\else$\nu_{\rm minL}$\fi}}
\newcommand{\Dnu}{\mbox{$\Delta \nu$}}
\newcommand{\muHz}{\mbox{$\mu$Hz}}
\newcommand{\kepler}{\mbox{\it{Kepler}}}
\newcommand{\corot}{\mbox{\it{CoRoT}}}
\newcommand{\numaxS}{\mbox{$\nu_{\rm max \sun}$}}
\newcommand{\MS}{{\rm M}\ifmmode_{\sun}\else$_{\sun}$~\fi}
\newcommand{\RS}{{\rm R}\ifmmode_{\sun}\else$_{\sun}$~\fi}
\newcommand{\LS}{{\rm L}\ifmmode_{\sun}\else$_{\sun}$~\fi}
\newcommand{\MSbit}{{\rm M}\ifmmode_{\sun}\else$_{\sun}$\fi}
\newcommand{\RSbit}{{\rm R}\ifmmode_{\sun}\else$_{\sun}$\fi}
\newcommand{\LSbit}{{\rm L}\ifmmode_{\sun}\else$_{\sun}$\fi}
\maketitle
\label{firstpage}
\begin{abstract}
%
Recently, by analysing the oscillation frequencies of 90 stars, Y\i ld\i z, \c{C}elik Orhan \& Kayhan have shown that the reference frequencies (\numin$_0$, \numin$_1$ and \numin$_2$) derived from glitches due to 
He {\scriptsize II} ionization zone have very 
strong diagnostic potential for the determination of their effective temperatures. In this study, we continue to analyse the same stars 
and compute their mass, radius and age from different scaling relations including relations based on \numin$_0$, \numin$_1$ and \numin$_2$.
For most of the stars, the masses computed using \numin$_0$ and  \numin$_1$ are very close to each other.
For 38 stars, the difference between these masses is less than 0.024 \MSbit. The radii of these stars from \numin$_0$ and  \numin$_1$ are even closer, 
with differences of less than 0.007 \RSbit.
These stars may be the most well known solar-like oscillating stars and deserve to be studied in detail.
The asteroseismic expressions we derive for mass and radius show slight dependence on metallicity. We therefore develop a new method for computing initial metallicity
from this surface metallicity by taking into account the effect of microscopic diffusion. 
The time dependence of initial metallicity shows some very interesting features that may be important for our understanding of chemical enrichment of Galactic Disc.
According to our findings, every epoch of the disc has its own lowest and highest values for metallicity. 
It seems that rotational velocity is inversely proportional to 1/2 power of age as given by the Skumanich relation.
 
\end{abstract}

\begin{keywords}
stars: evolution -- stars: fundamental parameters -- stars: interiors -- stars: late-type -- stars: oscillations.
\end{keywords}

\section{Introduction}
Determining the age of cool stars
is crucial for many branches of astrophysics, and precise determination of age from stellar properties primarily depends on how accurately the masses of such stars 
are calculated. Uncertainty of 10 per cent in mass corresponds to uncertainty of at least 30 per cent in age. 
The mass ($M$) and radius ($R$) of stars can be obtained by asteroseismic methods from the frequency of maximum amplitude (\numax), mean value of the large separation between oscillation 
frequencies ($\braket{\Dnu}$), and 
effective temperature (\teff). The relations used in these methods are called scaling relations. 
In Y\i ld\i z, \c{C}elik Orhan \& Kayhan  (2016; hereafter Paper III), new scaling relations were obtained based on variation 
of the first adiabatic exponent at stellar surface ($\Gamma_{\rm \negthinspace 1s}$).
However, for precise $M$ and $R$ determination, new methods are required because, in most cases, in particular \numax$ $ cannot be 
found very accurately from the power spectrum. Even for the Sun, the uncertainty in \numax$ $ ($\Delta$\numax) is not small enough. 
In the literature,
the range of solar \numax~ is 3021-3150 $\mu$Hz (see e.g. Stello et al. 2008; Kallinger et al. 2010; Huber et al. 2011; Mathur et al. 2012). 
The difference between the maximum and the miniumum of the literature values is about 4 per cent.
Since mass in scaling relation ($M_{\rm sca}$) is proportional to $\nu^3_{\rm max}$, the uncertainty in mass of such a star 
is very high, about 12 per cent. The uncertainty in frequencies of the minimum \Dnu, however, is much less than $\Delta$\numax.
Therefore, in order to obtain more accurate stellar parameters, we want to derive new scaling relations from the {\small MESA} models (Paper III) 
using the new reference frequencies found from the minimum \Dnu$ $ (\numin$_0$, \numin$_1$ and \numin$_2$; Y\i ld\i z et al. 2014a; hereafter Paper I).

The oscillatory component of frequency spacings is shaped by the He {\scriptsize II} ionization zone just below the stellar surface
(see e.g. fig. 7 in Paper I). 
Many quantities in such outer regions have a very sharp gradient and may be model dependent.
However, excellent results have already been obtained for the solar \teff, and for $M$ and $R$ of Procyon A using 
the new asteroseismic relations (Paper III). 
These successful applications motivate us to go further.

Two minima are clearly seen in the \Dnu$ $ versus $\nu$  graph for the Sun and solar model (figs 1-2 in Paper I).
We call the deepest minimum as the first minimum (min1) with $\numin_1$=2555.2 $\mu$Hz and the minimum with lower frequency
as the second minimum (min2), which is shallower than min1 for the Sun. For the main-sequence (MS) models with mass higher than the solar mass,
min2 becomes deeper and frequencies of the minima shifts toward low frequencies. There are also two minima in \Dnu$ $ versus $\nu$ graph
for the observed oscillation frequencies of the \kepler$ $ and \corot$ $ targets. From the comparison of observed and model frequencies of the 
minima we confirm that the minima with the lower frequency in general match min1. Therefore, we named the minima with higher frequency
as min0. For some of the stars, it is likely possible to obtain even min-1. 
In computation of \numin, we first determine
frequency interval of the minimum and draw two lines from the neighbourhood
intervals. The intersection of the two lines gives us the corresponding \numin.

In Paper III, we computed the effective temperature of stars from purely asteroseismic relations using 
\numin$_0$, \numin$_1$ and \numin$_2$. There are very clear relations between \teff$ $
and the order difference $\Delta n_{\rm xi} =(\numax-\numin_{\rm i})/\Dnu$.
For some stars, there are systematic differences between asteroseismic and observational \teff s
(spectroscopic and photometric). In such cases, we can modify the value of \numax$ $ so that all three quantities
\teff, $M_{\rm sca}$, and radius from scaling relations ($R_{\rm sca}$) are in good agreement with the values obtained in other ways (asteroseismic and
non-asteroseismic, see Section 4).
 
In this study, we analyse observed oscillation frequencies, check the relations between
the reference frequencies, and apply the new methods to 
the $Kepler$ and $CoRoT$ target stars. We find their $M$, 
\teff, $R$, luminosity ($L$), age ($t_{\rm sis}$),
and distance ($d_{\rm sis}$) using asteroseismic parameters. 
The role of the small separation between oscillation frequencies ($\delta \nu_{02}$) 
is crucial in the computation of $t_{\rm sis}$ 
for MS stars. Its mean value ($\braket{\delta \nu_{02}}$) is about 15 $\mu$Hz for zero-age 
MS (ZAMS) stars and 5 $\mu$Hz for terminal-age MS (TAMS) stars.
The distance ($d_{\rm obs}$) is also computed from very precise {\it Gaia} DR2 parallax (Gaia Collaboration 2018).

This paper is organized as follows. 
In Section 2, we develop new asteroseismic methods for 
the computation of $R$, surface gravity ($g$), $M$, and $t_{\rm sis}$ in terms of asteroseismic quantities.  
Some of these methods slightly depend on metallicity ($Z$). Therefore,
Section 3 is devoted to finding initial metallicity ($Z_{\rm o}$) from the present surface metallicity ($Z_{\rm s}$) by taking into account 
the effect of microscopic diffusion. 
In Section 4, we present the results obtained from applications of the methods developed in Sections 2 and 3 to the $Kepler$ and $CoRoT$ targets, and 
to some stars observed by ground-based telescopes.
The consequences of our findings in regards to the chemical evolution of Galactic Disc
and gyrochronology, {and comparison of asteroseismic and {\it Gaia} parallaxes} are presented in Section 5.
Finally, we draw our conclusions in Section 6.

\section{New scaling relations from the interior models using the {\small MESA} code} 
The models used in the analysis of this study are the same as those in Paper III. 
The details of the models constructed by using the {\small MESA} code (Paxton et al. 2011; Paxton et al. 2013) are given there. 
The symbols used in this study have the same meaning as in Paper III.

\subsection{Scaling relations in terms of $\nu_{\rm min1}$, $\braket{\Dnu}$, and $T_{\rm eff}$}
In some cases, \numax$ $ either cannot be determined from the observed data or its uncertainty is not low enough for accurate determination of stellar $M$ and $R$.
Therefore, using the {\small MESA} models with solar (initial) metallicity ($Z_\odot=0.0172$),
we also derive new relations using $\nu_{\rm min0}$, $\nu_{\rm min1}$ and $\nu_{\rm min2}$ in place of \numax.  
For radius ($R_{\rm sis1}$) in terms of $\nu_{\rm min1}$, $\braket{\Dnu}$ and $T_{\rm eff}$, we derive the fitting formula as 
\begin{equation}
\frac{R_{\rm sis1}}{R_{\sun}}=\frac{\left(\frac{\nu_{\rm min1}}{\nu_{\rm min1\odot}}\right)^{0.156}\left(\frac{\braket{\Delta \nu_\odot}}{\braket{\Dnu}}\right)^{0.92}}
{(1.14(r_{T\Gamma}-1.11)^2+0.98)(-0.64r_{\delta\Delta}+1.05)},
\label{equ:Rmin}
\end{equation}
where 
\begin{equation}
r_{T\Gamma}=\frac{T_{\rm eff}}{T_{\rm eff\odot}}\frac{\Gamma_{{\rm \negthinspace 1s}\odot}}{\Gamma_{\rm \negthinspace 1s}},
\end{equation}
and
\begin{equation}
r_{\delta\Delta}=\frac{\braket{\delta \nu_{02}}} {\braket{\Dnu}}.
\end{equation}
$R_{\rm sis1}$ is plotted \wrt model radius ($R_{\rm mod}$) in Fig. 1.
The maximum difference between equation (1) and $R_{\rm mod}$ is  about 1 per cent.
\begin{figure}
\includegraphics[width=101mm,angle=0]{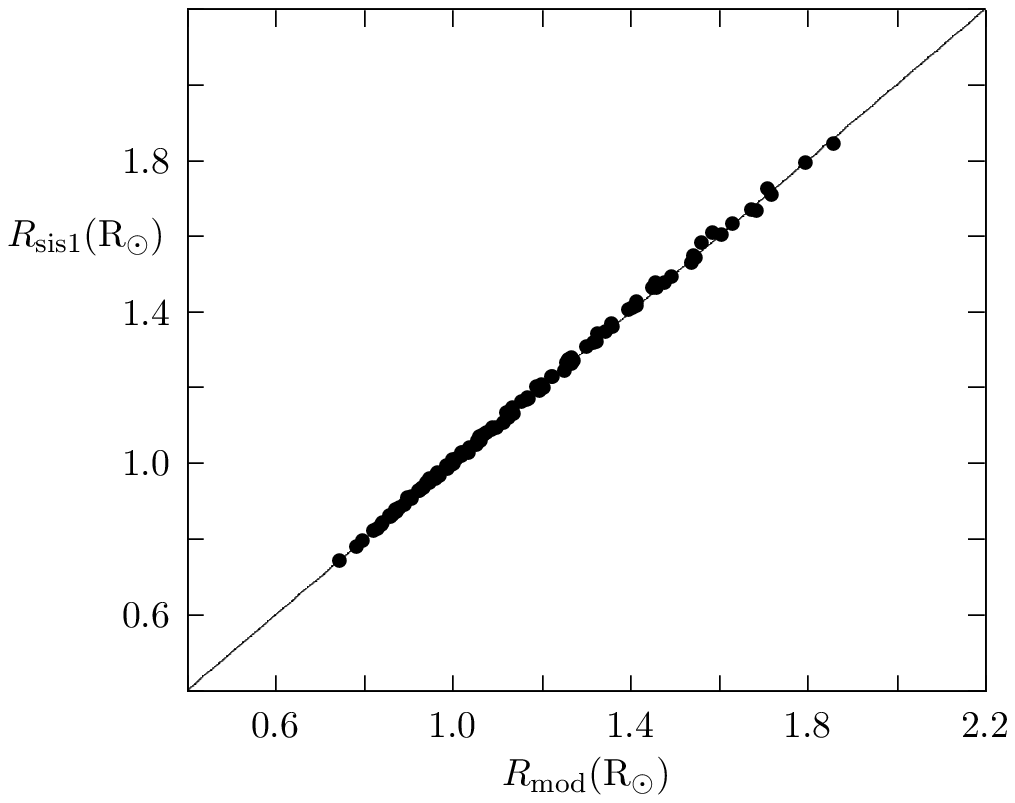}
\caption{Radius computed from asteroseismic data (equation 1) is plotted \wrt model radius (filled circles). 
The maximum difference between them is about 1 per cent.
}
\end{figure}

Similarly, we also derive an expression for 
asteroseismic gravity ($g_{\rm sis1}$) for the solar metallicity 
\begin{equation}
\frac{g_{\rm sis1}}{g_{\sun}}=\frac{\left(\frac{\nu_{\rm min1}}{\nu_{\rm min1\odot}}\frac{\braket{\Dnu}}{\braket{\Delta \nu_\odot}} \right)^{0.58}}
{(1.6(r_{T\Gamma}-1.06)^2+0.992)}.
\end{equation}
$g_{\rm sis1}$ is plotted \wrt model gravity ($g_{\rm mod}$) in Fig. 2.
The maximum difference between equation (4) and model gravity is about 2 per cent.

\begin{figure}
\includegraphics[width=101mm,angle=0]{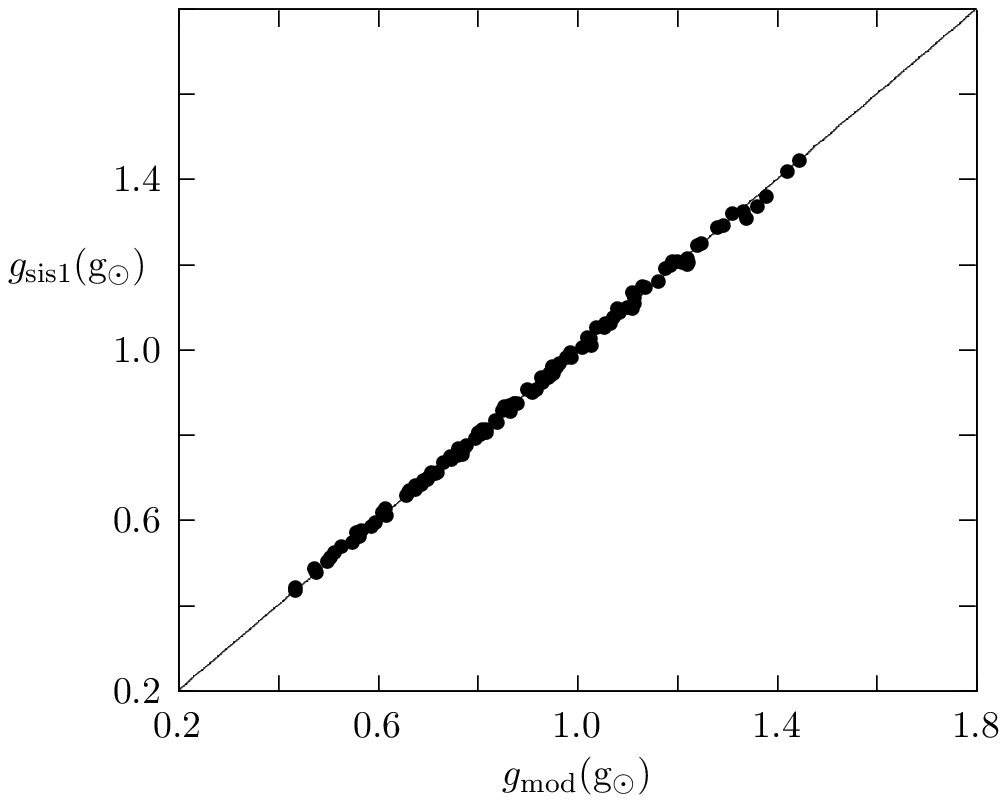}
\caption{Gravity at the surface computed from asteroseismic data (equation 4) is plotted \wrt model gravity. The fractional difference between these gravities 
is in between 0.02 and -0.02 for most of the stars. This implies that typical uncertainty is about 2 per cent.
}
\end{figure}

For stellar mass from min1, we compute $M_{\rm sis1}$ from $g_{\rm sis1}$ and $R_{\rm sis1}$: $M_{\rm sis1}/M_{\odot}= (g_{\rm sis1}/g_{\odot}) (R_{\rm sis1}/R_{\odot})^2$.
From the {\small MESA} models, there is a relation between $\nu_{\rm min0}$ and $\nu_{\rm min1}$:
\begin{equation}
\left(\frac{\nu_{\rm min1}}{\nu_{\rm min1\odot}}\right)=\left(\frac{\nu_{\rm min0}}{\nu_{\rm min0\odot}}\right)^{1.042}.
\end{equation}
By inserting equation (5) in equations (1) and (4),  we obtain expressions for radius ($R_{\rm sis0}$) and gravity ($g_{\rm sis0}$) 
in terms of $\nu_{\rm min0}$, respectively. Again, we follow the same steps for computation of $M_{\rm sis0}$ as for $M_{\rm sis1}$.

{One can ask if the surface effect on high-frequency modes may affect in particular value of $\nu_{\rm min0}$.
The effect can be illuminated by comparing relationships between  ${\nu_{\rm min1}}$ and $\nu_{\rm min0}$ of models and observations. 
The relationship between ${\nu_{\rm min1}}$ and $\nu_{\rm min0}$ given in equation (5) slightly changes if we use the observed frequencies.
The power of $\nu_{\rm min0}$ becomes 1.039. This implies that the surface effect in particular on $\nu_{\rm min0}$ is negligibly small.
The systematic difference is about 0.3 per cent.
}

$\Gamma_{\rm \negthinspace 1s\odot}$ is taken as 1.639.
The other solar quantities used in our methods are given at the end of Table B1.

\subsection{Effects of metallicity on the scaling relations}
Models with different metallicities show that there
is a slight dependence on $Z$, which can be important for a precise determination of $R$ and other basic stellar parameters.
We obtain
\begin{equation}
\frac{R_{\rm sis1}(Z)}{R_{\rm sis1}(Z=Z_\odot)}=0.09\left(\frac{Z}{Z_\odot}\right)^{0.8}+0.903.      
\end{equation}
For gravity with arbitrary $Z$,
\begin{equation}
\frac{g_{\rm sis1}(Z)}{g_{\rm sis1}(Z=Z_\odot)}=0.026\left(\frac{Z}{Z_\odot}\right)+0.974      
\end{equation}
is derived. Indeed, equations (6) and (7) show that both $R_{\rm sis1}$ and $g_{\rm sis1}$ slightly depend
on $Z$, respectively. For example, if $Z=2Z_\odot$, $R_{\rm sis1}(Z)$ and $g_{\rm sis1}(Z)$ 
are only 4 and 2 per cent greater than their values with solar metallicity, respectively. However,  
as long as we deal with fundamental parameters such as mass and radius, such a level of precision 
is important.

We do the same computations for min0 and obtain $g_{\rm sis0}(Z)$, $R_{\rm sis0}(Z)$ and $M_{\rm sis0}(Z)$ (see Table B1).

\subsection{Computation of age }
\begin{figure}
\includegraphics[width=97mm,angle=0]{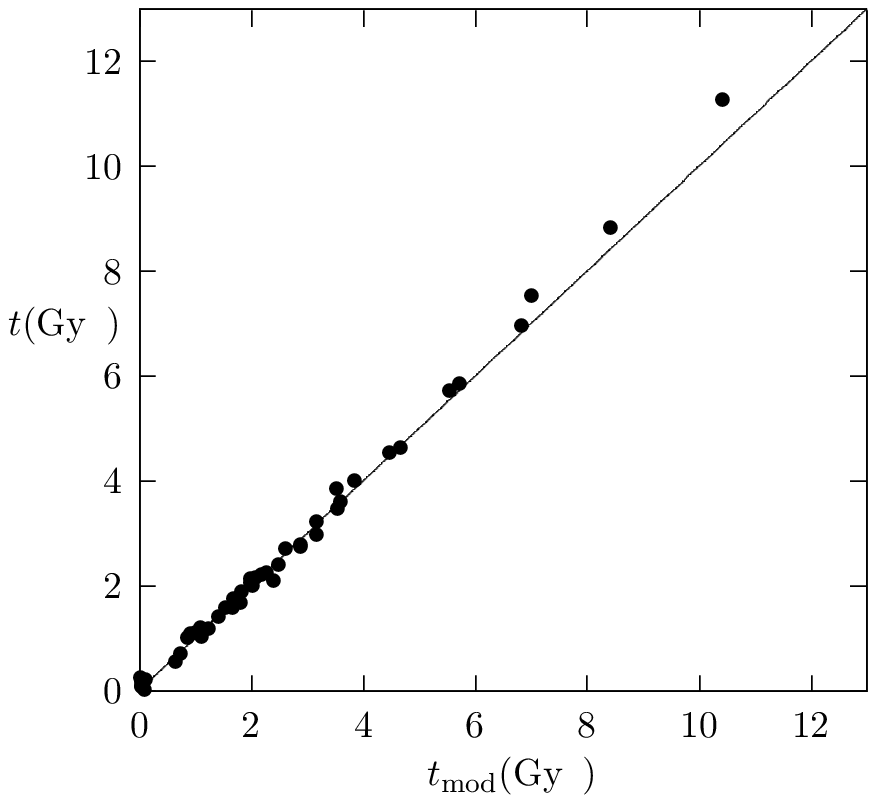}
\caption{Age derived from oscillation frequencies (equation 8) with
respect to model age. Mass range is 0.9-1.6 \MSbit. The agreement between 
the two ages is very good, particularly for ages less than 7 Gyr.
}
\end{figure}
{
Age of an MS star is uncertain because it could either be a ZAMS or a TAMS star, and ZAMS and TAMS ages are totally different. 
Therefore, $\braket{\delta \nu_{02}}$ works very well in computation of accurate age of MS stars.   
For a post-MS star, however, its age is close to its TAMS age. 
}

For the {\small MESA} models with mass ranging from 0.9 to 1.6 \MSbit, we derive a fitting formula for $t_{\rm sis}$ in terms of $\braket{\delta \nu_{02}}$,
$M$ and $Z$ as 
\begin{equation} 
t_{\rm sis}({\rm Gyr})= \frac{a_t(Z)\braket{\delta \nu_{02}}/\braket{\delta \nu_{02\sun}}(M/\MS)^{1.2}+b_t(Z)}{(M/\MS)^{n_{t}(Z)}}
\end{equation}
where 
\begin{equation} 
a_t(Z)=9.84 (0.931\left(\frac{Z_\odot}{Z}-0.331\right)^{0.2}-1.627),
\end{equation}
\begin{equation} 
b_{t}(Z)=6.176\left(\frac{Z}{Z_\odot}\right)+6.016
\end{equation}
and
\begin{equation} 
n_{t}(Z)=-0.2447\left(\frac{Z_\odot}{Z}\right)+3.3848.
\end{equation}
$n_{t}(Z)$ is the slope obtained from the graph of $\log(t)$ \wrt $\log(M)$.

$\braket{\delta \nu_{02}}$ (or $\braket{\delta \nu_{02}}/\braket{\delta \nu_{02\sun}}$) is considered
the best age indicator of a star; however, it actually represents relative age in MS. 
Relative age can be defined as the age in units of TAMS age, which strongly depends on stellar mass, 
and metallicity. Therefore, in order to find absolute age, stellar mass and metallicity must be precisely obtained from 
observations.
$t_{\rm sis}$ found from equation (8) is plotted \wrt model age in Fig. 3. They are in very good agreement, especially for 
ages less than 7 Gyr.

We also compute age using the method developed by Y\i ld\i z et al. (2014b) for the planet-candidate host stars. This method is based on
the observed $M$, $R$ and $Z$ values of a star. This age ($t_{\rm yil}$) and $t_{\rm sis}$ from equation (8) are listed in Table B1.

\subsection{Uncertainties in radius, gravity, mass, and age  }
From equation (1), the typical uncertainty ($\Delta R_{\rm sis1}$) in $R_{\rm sis1}$ is derived as
\begin{equation}
\frac{\Delta R_{\rm sis1}}{R_{\rm sis1}}=
0.156\frac{\Delta \numin_1}{\numin_1} 
+0.92\frac{\Delta \braket{\Delta \nu}}{\braket{\Delta \nu}}.
\end{equation}
Similarly, uncertainty in gravity ($\Delta g_{\rm sis1}$)  is obtained as 
\begin{equation}
\frac{\Delta g_{\rm sis1}}{g_{\rm sis1}}=
0.58\frac{\Delta \numin_1}{\numin_1} 
+0.58\frac{\Delta \braket{\Delta \nu}}{\braket{\Delta \nu}}
\end{equation}
using equation (4).

Since we compute mass by multiplying the square of equation (1) by equation (4),
it can be shown that the typical uncertainty in mass is given as
\begin{equation}
\frac{\Delta M_{\rm sis1}}{M_{\rm sis1}}=
0.89\frac{\Delta \numin_1}{\numin_1} 
+1.26\frac{\Delta \braket{\Delta \nu}}{\braket{\Delta \nu}}.
\end{equation}

For $M_{\rm sis0}$, $R_{\rm sis0}$, and $g_{\rm sis0}$, we use the same methods but 
${\Delta \numin_0}/{\numin_0}$ is employed in place of ${\Delta \numin_1}/{\numin_1}$.

\section{Computation of initial metallicity from present surface metallicity and microscopic diffusion}
From comparison of solar models with the helioseismic inferences, such as sound speed, bottom radius of the convective zone (CZ),
and surface helium abundance, one can expect that microscopic diffusion works throughout the radiative interior and ultimately affects 
the composition of the solar CZ and photosphere (Michaud \& Proffitt 1993; Thoul, Bahcall \& Loeb 1994; Bahcall, Serenelli \& Pinsonneault 2004; \yildiz 2011). Further confirmation can perhaps be carried out in clusters, particularly old open clusters.
In at least two clusters, there are strong indicators that diffusion works (see below).

It is well known that rotation causes mixing and therefore prevents diffusion of chemical species. MS stars with convective envelopes
are slow rotators and therefore one can observe the influence of diffusion in these stars. However, the efficiency of diffusion also 
depends on the depth of CZ. The shallower the CZ is, the faster the diffusion velocity at the bottom of CZ is. Therefore,
there must be a metallicity gradient along the MS of an old cluster. Metallicity is minimum at the turn-off (TO) and increases toward the cool side of MS.
As stars evolve away from MS, CZ deepens and mixes the outer and inner regions. Therefore, near the red giant branch (RGB), the metallicity at 
the surface is approximately the same as initial metallicity. The metallicity difference between RGB and TO is found as 
0.08 dex for NGC 6752 (Gruyters, Nordlander \& Korn  2014), 
0.16 dex for NGC 6397 (Korn et al. 2007),   
0.25 dex for M30 (Gruyters et al. 2016), 
and  
0.26 dex for M92 (King et al. 1998). 

\subsection{Surface and initial metallicities}
We have computed metallicity of the stars from the [Fe/H]  values derived from spectroscopic studies (Bruntt et al. 2012; Molenda{-\.{Z}}akowicz et al. 2013). As in the case of \yildiz et al. (2014b),
we compute [O/H] value using the relation between [O/H] and [Fe/H] abundances in the solar neighbourhood (Edvardsson et al. 1993) and obtain Z from the relation
\begin{equation} 
Z=10^{\rm [O/H]} Z_{\sun}.
\end{equation} 
This metallicity is the metallicity in the photospheres of the stars ($Z_{\rm s}$). However, 
we want to find 
the initial metallicity ($Z_{\rm o}$).
   
Since metallicity has a slight effect on the scaling relations for radius and gravity (and hence mass), the initial metallicity must be estimated from present surface value and
microscopic diffusion velocity.
\subsection{Effect of diffusion on surface metallicity of MS stars}
The surface metallicity and helium abundance of a slowly rotating cool star decrease during its MS evolution. 
The difference ($\delta Z$) between $Z_{\rm o}$ and $Z_{\rm s}$ is a function of both age and mass of stars.
From the {\small ANK\.I} models of  Y\i ld\i z, \c{C}elik Orhan \& Kayhan (2015; hereafter Paper II) with different mass and metallicity, we find the difference  ($\delta Z$) between $Z_{\rm o}$ and $Z_{\rm s}$ as
\begin{equation} 
\delta Z=Z_{\rm o}-Z_{\rm s}=5.69\times 10^{-4}((M/\MS)^{6.73}+0.7)t_9^{0.9}
\end{equation} 
where $t_9$ is the age in unit of $10^9$ yr.
Microscopic diffusion causes a decrease in the photospheric metallicity of a star relative to its age and mass.

\subsection{Effect of deepening CZ on surface metallicity of post-MS stars}
The surface metallicity of a cool star increases after its MS phase 
because deepening CZ mixes low-metallicity outer regions with the metal-rich inner regions. Such a differentiation in metallicity is 
thought to be caused by microscopic diffusion. 
For post-MS stars, we introduce a parameter as 
\begin{equation} 
x_{gT}=\frac{\teff}{\teffsun}\log g -3.35.
\end{equation} 
If $x_{gT}< 0$, then $Z_{\rm o}=Z_{\rm s}/1.06$. If $x_{gT}> 0$,
\begin{equation} 
Z_{\rm o}=\frac{Z_{\rm s}}{1-0.526 x_{gT}}.
\end{equation}

\section{Results and Discussions}
The basic properties of the \kepler$ $ and \corot$ $ targets are listed in Table B1. 
The observational properties we use 
are the same as those given in Paper III. 
In this study we compute fundamental properties using methods based in particular on $\numin_0$ and $\numin_1$
rather than \numax,
because \numax$ $ is not always precisely determined from the power spectrum (Arentoft et al. 2008). 
If the parameters obtained by using  $\numin_0$ and $\numin_1$ are consistent, then
we can slightly 
modify only \numax$ $ by comparing values of \teff, $M$, and $R$ found by using different methods (see Section 4.2). 

\subsection{Mass, radius, distance, and age of the target stars}

In Fig. 4, $R_{\rm sis1}$ computed from min1 (equations 1 and 6) is plotted \wrt $R_{\rm sis0}$ from min0. The agreement between 
these radii seems excellent. The difference between $R_{\rm sis1}$ and $R_{\rm sis0}$ is less than 1 per cent.
Such a level of accuracy is achieved for the first time and reflects the diagnostic potential of new reference frequencies
$\nu_{\rm min0}$, $\nu_{\rm min1}$, and $\nu_{\rm min2}$. 

For $M_{\rm sis1}$, we first compute $g_{\rm sis1}$ from equation (4) and (7) and then multiply it by $R^2_{\rm sis1}$. 
In a similar way, we obtain $M_{\rm sis0}$ using $\numin_0$ in place of $\numin_1$.
In Fig. 5, $M_{\rm sis1}$ is plotted \wrt $M_{\rm sis0}$. The agreement between
these masses is amazing. The maximum difference between them is about 4 per cent. However, for 38 stars     
(including the Sun), the mass difference ($\delta M_{01}$) between
$M_{\rm sis1}$ and $M_{\rm sis0}$ is less than 0.024 \MSbit. These stars are listed in Table 1. 
The difference between $R_{\rm sis1}$ and $R_{\rm sis0}$ ($\delta R_{01}$) is given in the ninth column.
The maximum value of $\delta R_{01}$ is about 0.007 \RS for KIC 11717120, the percentage difference is  0.3.
These solar-like oscillating stars may be the most well known stars.
Their interior models must be studied in detail.
Two of these stars, KIC 7871531 and KIC 8760414, have mass low enough to discuss non-ideal effects in the stellar interior. Their masses are calculated as 0.80 and 0.85 \MSbit, respectively. 

The masses of two stars are already well known from non-asteroseismic observational methods. They are the Sun and Procyon A.
Both $M_{\rm sis1}$ and $M_{\rm sis0}$ are 0.99 \MS for the Sun. For Procyon A,
$M_{\rm sis1}=1.45$ \MS while its mass was determined to be 1.478 \MS $\pm$ 0.012 from the astrometric data 
of the {\it Hubble Space Telescope} (Bond et al. 2015). These results show that the masses found by using the new reference frequencies 
$\nu_{\rm min0}$ and $\nu_{\rm min1}$ are very accurate.

Much more precise values are derived for radius. For the Sun, for e.g. both $R_{\rm sis1}$ and $R_{\rm sis0}$ are 1.00 \RSbit.
There is also satisfactory agreement between the radii of Procyon A determined using asteroseismic and non-asteroseismic methods (see below).
\begin{table*}
\caption{ The most precise data for the solar-like oscillating stars: 
$\delta M_{\rm 01}= \mid M_{\rm sis0}- M_{\rm sis1}\mid <0.024$ \MSbit and $\delta R_{\rm 01}= \mid R_{\rm sis0}- R_{\rm sis1}\mid< 0.007$ \RSbit.
For Procyon A, $\delta M$ is the difference between $ M_{\rm sis1}$ and $ M_{\rm sca}$.
}
\centering
\begin{tabular}{rlllclllccc}

\hline
Star & $M_{\rm sca}$ & $ M_{\rm sis0}$ &$ M_{\rm sis1}$ & $\delta M_{01}$ & $R_{\rm sca}$ &$ R_{\rm sis0}$ &$ R_{\rm sis1}$ & $\delta R_{01}$  & $Z_{\rm s}$ & $Z_{\rm o}$ \\
        & (\MS)           &  (\MS)  &  (\MS)     &  (\MS)   & (\RS)     &  (\RS)  &  (\RS)     &  (\RS)   &    &     \\
\hline
{ 3427720} &     1.09 &     1.11 &     1.09 &     0.024 &     1.11 &     1.11 &     1.11 &     0.004 &   0.0115 &  0.0140  \\
{ 3544595} &     0.90 &     0.88 &     0.90 &     0.021 &     0.92 &     0.91 &     0.92 &     0.004 &   0.0100 &  0.0136  \\
{ 3632418} &     1.24 &     1.24 &     1.26 &     0.019 &     1.83 &     1.83 &     1.83 &     0.005 &   0.0095 &  0.0152  \\
{ 5184732} &     1.17 &     1.17 &     1.18 &     0.013 &     1.34 &     1.34 &     1.34 &     0.003 &   0.0184 &  0.0259  \\
{ 5607242} &     1.09 &     1.08 &     1.07 &     0.011 &     2.31 &     2.33 &     2.33 &     0.004 &   0.0111 &  0.0125  \\
{ 5866724} &     1.26 &     1.25 &     1.25 &     0.000 &     1.41 &     1.42 &     1.42 &     0.000 &   0.0145 &  0.0212  \\
{ 5955122} &     1.25 &     1.20 &     1.21 &     0.012 &     2.11 &     2.12 &     2.13 &     0.004 &   0.0112 &  0.0171  \\
{ 6116048} &     1.01 &     1.00 &     1.01 &     0.005 &     1.22 &     1.22 &     1.22 &     0.001 &   0.0089 &  0.0131  \\
{ 6933899} &     1.14 &     1.14 &     1.15 &     0.007 &     1.60 &     1.61 &     1.61 &     0.002 &   0.0125 &  0.0200  \\
{ 7206837} &     1.38 &     1.39 &     1.38 &     0.009 &     1.59 &     1.59 &     1.59 &     0.002 &   0.0140 &  0.0224  \\
{ 7747078} &     1.12 &     1.10 &     1.12 &     0.019 &     1.94 &     1.95 &     1.95 &     0.006 &   0.0087 &  0.0128  \\
{ 7871531} &     0.81 &     0.80 &     0.80 &     0.004 &     0.87 &     0.87 &     0.87 &     0.001 &   0.0089 &  0.0126  \\
{ 8228742} &     1.24 &     1.24 &     1.23 &     0.004 &     1.81 &     1.81 &     1.81 &     0.001 &   0.0101 &  0.0161  \\
{ 8379927} &     1.09 &     1.09 &     1.10 &     0.014 &     1.11 &     1.11 &     1.11 &     0.003 &   0.0115 &  0.0139  \\
{ 8524425} &     1.10 &     1.07 &     1.05 &     0.017 &     1.79 &     1.81 &     1.81 &     0.005 &   0.0140 &  0.0193  \\
{ 8694723} &     1.10 &     1.10 &     1.09 &     0.004 &     1.52 &     1.53 &     1.53 &     0.001 &   0.0071 &  0.0114  \\
{ 8760414} &     0.85 &     0.85 &     0.85 &     0.001 &     1.04 &     1.03 &     1.03 &     0.000 &   0.0037 &  0.0054  \\
{ 9025370} &     0.97 &     0.97 &     0.97 &     0.002 &     1.00 &     0.99 &     0.99 &     0.000 &   0.0075 &  0.0107  \\
{ 9098294} &     0.97 &     0.97 &     0.97 &     0.006 &     1.15 &     1.14 &     1.14 &     0.001 &   0.0092 &  0.0134  \\
{ 9139151} &     1.15 &     1.16 &     1.14 &     0.024 &     1.15 &     1.15 &     1.15 &     0.004 &   0.0136 &  0.0163  \\
{ 9410862} &     1.01 &     1.01 &     1.01 &     0.006 &     1.17 &     1.17 &     1.17 &     0.001 &   0.0078 &  0.0114  \\
{ 9812850} &     1.25 &     1.26 &     1.25 &     0.013 &     1.74 &     1.74 &     1.74 &     0.003 &   0.0092 &  0.0146  \\
{10018963} &     1.24 &     1.23 &     1.21 &     0.015 &     1.95 &     1.96 &     1.96 &     0.004 &   0.0086 &  0.0138  \\
{10454113} &     1.16 &     1.16 &     1.16 &     0.002 &     1.25 &     1.25 &     1.25 &     0.000 &   0.0111 &  0.0149  \\
{10920273} &     1.11 &     1.09 &     1.09 &     0.000 &     1.85 &     1.86 &     1.86 &     0.000 &   0.0116 &  0.0164  \\
{11026764} &     1.13 &     1.11 &     1.10 &     0.008 &     2.03 &     2.05 &     2.04 &     0.003 &   0.0126 &  0.0167  \\
{11244118} &     1.21 &     1.20 &     1.21 &     0.009 &     1.64 &     1.65 &     1.65 &     0.002 &   0.0156 &  0.0249  \\
{11253226} &     1.35 &     1.35 &     1.37 &     0.019 &     1.59 &     1.56 &     1.57 &     0.004 &   0.0094 &  0.0150  \\
{11295426} &     1.05 &     1.02 &     1.05 &     0.022 &     1.24 &     1.23 &     1.24 &     0.005 &   0.0137 &  0.0194  \\
{11414712} &     1.08 &     1.07 &     1.08 &     0.016 &     2.18 &     2.22 &     2.22 &     0.006 &   0.0112 &  0.0138  \\
{11717120} &     0.95 &     0.86 &     0.88 &     0.015 &     2.30 &     2.32 &     2.32 &     0.007 &   0.0083 &  0.0078  \\
{12009504} &     1.12 &     1.11 &     1.12 &     0.012 &     1.38 &     1.38 &     1.38 &     0.003 &   0.0101 &  0.0154  \\
{12069424} &     1.03 &     1.03 &     1.03 &     0.005 &     1.21 &     1.21 &     1.21 &     0.001 &   0.0135 &  0.0190  \\
{12258514} &     1.19 &     1.19 &     1.18 &     0.017 &     1.58 &     1.58 &     1.58 &     0.004 &   0.0125 &  0.0200  \\
{    2151} &     1.10 &     1.09 &     1.09 &     0.001 &     1.83 &     1.85 &     1.85 &     0.000 &   0.0106 &  0.0156  \\
{   43587} &     1.07 &     1.06 &     1.07 &     0.018 &     1.20 &     1.19 &     1.20 &     0.004 &   0.0116 &  0.0169  \\
{   49385} &     1.31 &     1.24 &     1.26 &     0.018 &     1.96 &     1.96 &     1.96 &     0.005 &   0.0132 &  0.0212  \\
{ Procyon A} &     1.47 &     --- &     1.45 &     0.014 &     2.03 &     --- &     2.01 &       --- &   0.0123 &  0.0197  \\
{$\odot $} &     1.00 &     0.99 &     0.99 &     0.000 &     1.00 &     1.00 &     1.00 &     0.000 &   0.0134 &  0.0167  \\
\hline
\end{tabular}
\end{table*}

\begin{figure}
\includegraphics[width=101mm,angle=0]{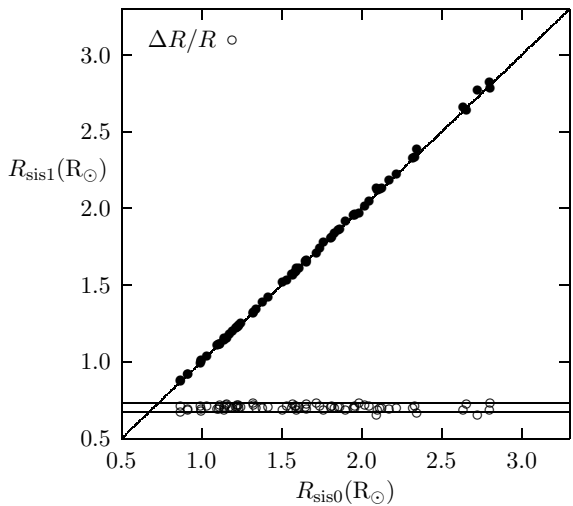}
\caption{$R_{\rm sis1}$ is plotted \wrt $R_{\rm sis0}$.
The horizontal lines represent 0.01 and -0.01 for the uncertainties in radius. 
}
\end{figure}
\begin{figure}
\includegraphics[width=101mm,angle=0]{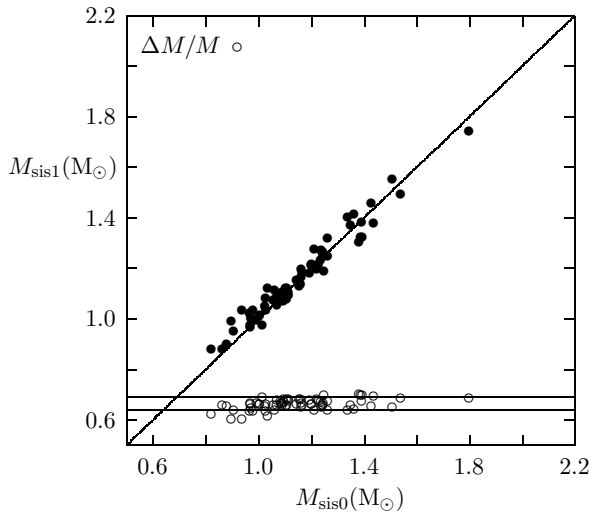}
\caption{$M_{\rm sis1}$ is plotted \wrt $M_{\rm sis0}$.
The horizontal lines represent 0.04 and -0.04 for the uncertainties in mass. 
}
\end{figure}

For most of the stars, both $d_{\rm sis}$ and $d_{\rm obs}$ are available. There is in general a good agreement between them
(see Section 5.3).
Detailed analysis of asteroseismic and non-asteroseismic distances of about 1800 stars is the subject of another study
(\yildiz \& \"Ortel, in preparation).
 
Age is computed from asteroseismic quantities using equation (8). 
Except KIC 7871531, the age of stars is less than 9 Gyr. We also compute age using the method derived by \yildiz et al. (2014b) 
for planet-candidate host stars. 
KIC 3733735 has very small age: $t_{\rm sis} =0.07 $ Gyr. It is a fast rotator (13 km\,s$^{-1}$). Its seismic radius is small according to its mass and $Z_{\rm o}$, less than ZAMS radius
(see Section 4.3). 


\subsection{Effective temperature difference between asteroseismic and non-asteroseismic methods 
and modification in \numax}
In Papers I and III,
it is shown that \teff$ $ is a function of order difference between the minima in \Dnu$ $ versus $\nu$ graph and
\numax. Using equations (15)-(17) of Paper III, \teff s of the stars are computed by using their $\nu_{\rm min0}$, $\nu_{\rm min1}$, and $\nu_{\rm min2}$ 
together with \numax$ $ and $\Dnu$.
New expressions are derived for $T_{\rm sis0}$, $T_{\rm sis1}$, and $T_{\rm sis2}$ by taking $f_\nu=1$ (see Appendix).
For most of the stars, \teff s are available from spectroscopic observations and also from colours ($B-V$ and $V-K$).
There is in general a very good agreement between these \teff s. However, in some cases there is a systematic difference between the 
\teff$ $ values derived from asteroseismic and non-asteroseismic methods. This can be caused by values of \numax,
which is perhaps the most uncertain asteroseismic quantity. Therefore, we slightly 
modify the \numax$ $ value to determine whether the systematic difference can be eliminated.
At the same time, we test if mass and radius computed from new scaling relations are improved 
in comparison with the values obtained using asteroseismic methods based on the frequencies of minimum 
$\Dnu$.

In many cases, the resultant $\teff$, $M_{\rm sca}$, and $R_{\rm sca}$ from the modified scaling relations (equations 9 and 10 in Paper III) come to close to 
the values ($M_{\rm sis1}$ and $R_{\rm sis1}$) from alternative scaling relations from minima (equations 1 and 4). 
In Figs 6 and 7, $ M_{\rm sis1}$ and $ R_{\rm sis1}$ are plotted 
\wrt $ M_{\rm sca}$ and $ R_{\rm sca}$, respectively.
The agreement between  $R_{\rm sis1}$ and $ R_{\rm sca}$ is in general excellent. This is the case for $M_{\rm sis1}$ and $ M_{\rm sca}$ also for most of the
stars. There is a significant difference between $M_{\rm sis1}$ and $ M_{\rm sca}$ for only KIC 8561221. 
When we compare $M_{\rm sis0}$ and $ M_{\rm sca}$, we confirm a few more scattering in the data. This may be due to shallow depth of min0,
which can increase the uncertainty in $\nu_{\rm min0}$.
\begin{figure}
\includegraphics[width=101mm,angle=0]{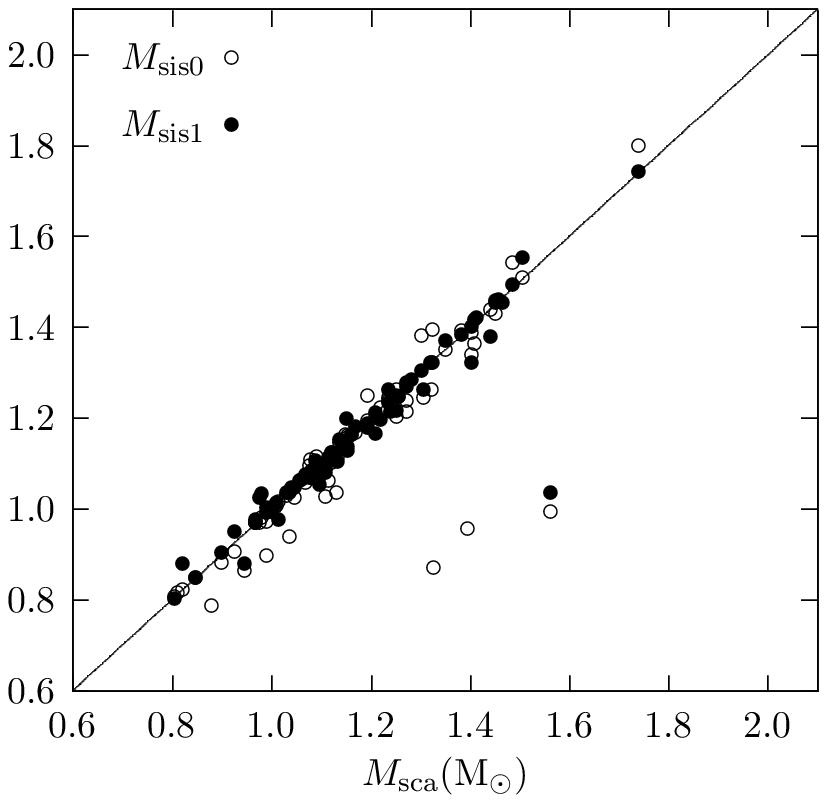}
\caption{$M_{\rm sis0}$ (circles) and $M_{\rm sis1}$ (filled circles)  are plotted \wrt $M_{\rm sca}$.
}
\end{figure}
\begin{figure}
\includegraphics[width=101mm,angle=0]{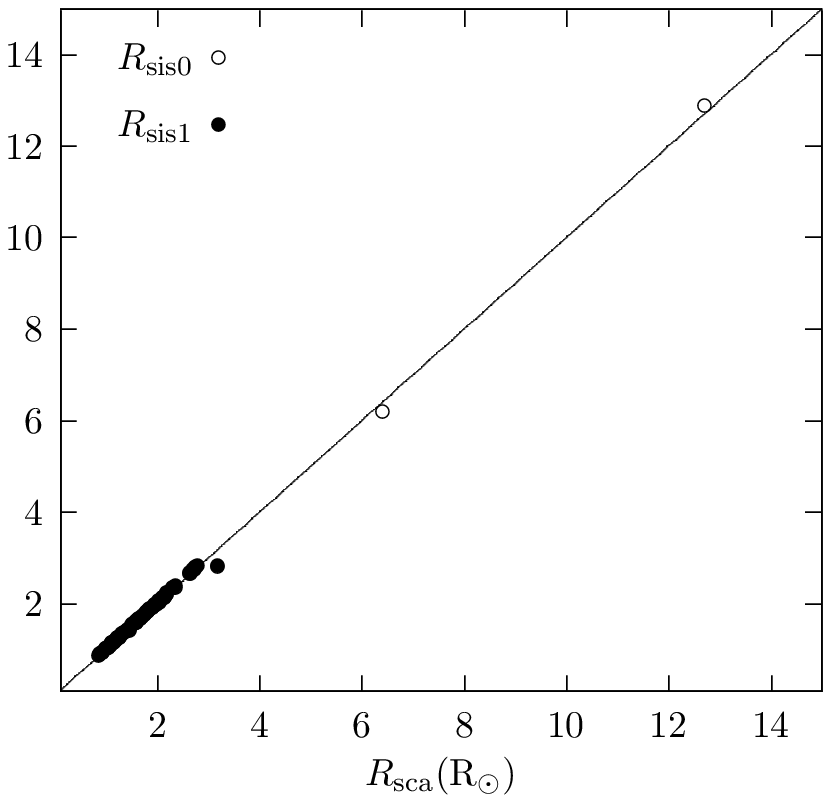}
\caption{$R_{\rm sis0}$ (circles) and $R_{\rm sis1}$ (filled circles)  are plotted \wrt $R_{\rm sca}$.
}
\end{figure}

\subsection{Notes on individual stars}
Comparison of asteroseismic and {\it Gaia} DR2 distances (Gaia Collaboration 2018) of individual stars is a useful way
to test the methods we have developed. 
Asteroseismic distance of the targets is computed by applying the same method as in \yildiz et al. (2017).
For computation of luminosity, we use radius found from scaling relation with modified \numax$ $ and $T_{\rm eS}$.  
The {\it Gaia} DR2 and asteroseismic distances are listed in Table B1. For 6 of the targets, {\it Gaia} parallax is not available. Their distances are
computed from 
{\it Hipparcos} (Perrymann et al. 1997) parallax (HD 2151,   7.46$\pm$0.01 pc;
HD 43587,  19.25$\pm$0.15 pc;
HD 203608,  9.26$\pm$0.02 pc;
Procyon A,    3.51$\pm$0.02 pc;
KIC 7747078  184$\pm$8 pc;
KIC 11026764 184$\pm$5 pc). 


\subsubsection{KIC 3424541}
KIC 3424541 is a sub-giant star. Two minima are seen on its \Dnu$ $ versus $\nu$ diagram. 
It shows mixed modes in its observed oscillation frequencies. 
Although there is a systematic difference between asteroseismic  and non-asteroseismic \teff s,
for such hot solar-like oscillating stars, asteroseismic \teff$ $ is only slightly dependent on the value of \numax.
However, if we take  $\numax=722$ $\mu$Hz, $R_{\rm sca}$ is in very good agreement with $R_{\rm sis0}$ and  $R_{\rm sis1}$,
and $M_{\rm sca}$  becomes  very close to  $M_{\rm sis1}$. 

\subsubsection{KIC 3427720}
Effective temperatures derived from min1 and min0 of KIC 3427720 are lower than both photometric and spectral $\teff$s if we use literature \numax.
Decreasing \numax$ $ may remove this systematic difference. If we take $\numax=2602$ $\mu$Hz, then 
one can obtain a very good agreement between asteroseismic and conventional  $\teff$s. Furthermore, the modified \numax$ $ makes $M_{\rm sca}$ and $R_{\rm sca}$
in perfect agreement with the literature values and the values derived from  $\nu_{\rm min0}$ and $\nu_{\rm min1}$. 

\subsubsection{KIC 3544595}
It is one of the brightest $Kepler$ targets and a planet host star with two planets (Ballard et al. 2014). If 
$\numax$ is taken as $3247$ $\mu$Hz, the masses, radii, and $\teff$s computed with different methods are in much better agreement in comparison to the 
case with the observed value of $\numax=3366$ $\mu$Hz.

\subsubsection{KIC 3632418} 
The F-like oscillating star KIC 3632418 has a rocky planet (Howell et al. 2012). On the \Dnu$ $ versus $\nu$ diagram of KIC 3632418, it is quite difficult to
decide on the minima. If we plot for degrees $l=0$, $l=1$, and $l=2$ on the \Dnu$ $ versus $\nu$ diagram of KIC 3632418, 
min1 is not confirmed clearly from the data of all degrees. For degree $l=0$, there is a bump-like structure in the vicinity of min1, 
and min0 is not seen due to lack of data. Therefore, we determine the min1 and min0 from $l=1$.
If we use the observed value of $\numax= $ 1159 $\mu$Hz  in scaling relations, asteroseismic $\teff$s are slightly different
from conventional $\teff$s. However, if we take $\numax= $ 1077 $\mu$Hz, the agreement between the radii from different methods is improved.
The differences between the radii, then, are less than 0.01 \RSbit. 

\subsubsection{KIC 3656476}
The observed oscillation frequencies of $Kepler$ target KIC 3656476  do not allow computation of the frequency of any minima. 
Therefore, we cannot obtain $T_{\rm sis0}$ and $T_{\rm sis1}$ for this star. 
\numax$ $ and \Dnu$ $ are available in the literature. 
We evaluate mass and radius using  the scaling relations. 
The differences $\Delta M = |M_{\rm sca}-M_{\rm lit}|$ and $\Delta R = |R_{\rm sca}-R_{\rm lit}|$ are small.

\subsubsection{KIC 3733735}
Its $T_{\rm eS}$ is in good agreement with its $T_{\rm eVK}$. 
The distance of KIC 3733735 is 102.8 pc and in good agreement with $d_{\rm sis}$. min0 cannot be determined from observed frequencies, but this star has min2 and min1. 
We can compute effective temperatures, masses, and radii from \numin$_1$ and \numin$_2$. 
If we decrease \numax$ $ by about 49 $\mu$Hz, $M_{\rm sca}$ and $R_{\rm sca}$ are in very good agreement with
$M_{\rm sis1}$ and $R_{\rm sis1}$, respectively.
Different references give different values for \numax$ $ and hence for its $M_{\rm sca}$ and $R_{\rm sca}$.

\subsubsection{KIC 4349452}
KIC 4349452 has three planets. Two of these planets are Neptune-sized (Benomar et al. 2014).
If we reduce the observed \numax$ $ by only 106 $\mu$Hz, $T_{\rm sis0}$ and $T_{\rm sis1}$ are in much better agreement with $T_{\rm eS}$.
$R_{\rm sca}$ with modified \numax$ $, $R_{\rm sis0}$ and $R_{\rm sis1}$ are very close to each other.

\subsubsection{KIC 4914923} 
KIC 4914923 is a high proper motion star. Neilsen et al. (2015) have identified rotational splitting of its p-mode frequencies.
The rotation period of KIC 4914923 is determined as $1.23 \pm 0.29$ in unit of solar rotation period.

The observed \numax$ $ of KIC 4914923 is given as 1849  $\mu$Hz.
From its oscillation frequencies, we find \numin$_0$ as 1947.8 $\mu$Hz. The difference between $T_{\rm eS}$  and $T_{\rm sis0}$ is about 173 K. 
While $R_{\rm sis0}$ and $R_{\rm lit}$ are as 1.38 and 1.37 \RSbit, this value of \numax$ $ yields $R_{\rm sca}=1.43$ \RSbit.
If we reduce $\numax$ to 1775 $\mu$Hz, the difference between $T_{\rm eS}$  and $T_{\rm sis0}$  becomes only 24 K and $R_{\rm sca}$ is found as 1.37 \RS, in much better
agreement with $R_{\rm sis0}$ and $R_{\rm lit}$. $M_{\rm sis0}$, $M_{\rm lit}$, and $M_{\rm sca}$ with modified $\numax$ are about 1.10 \MS.

\subsubsection{KIC 5184732}
\numin$_0$ and \numin$_1$ of KIC 5184732 are determined from observed oscillation frequencies of modes with $l=1$.
Half of min1 is seen in the \Dnu$ $ versus $\nu$ graph, which implies that the assigned value for \numin$_1$ (1705.5 $\mu$Hz) can be considered the upper limit.
Its $T_{\rm eVK}$ is in excellent agreement with $T_{\rm eS}$.  The observed value of \numax$ $ (2068 $\mu$Hz) gives $T_{\rm sis0}$ as 180 K less than $T_{\rm eS}$.
We notice that $M_{\rm sca}$ and $R_{\rm sca}$ are greater than $M_{\rm sis0}$ and $R_{\rm sis0}$, respectively. These systematic differences can be eliminated
by modifying the value of \numax. If \numax$ $ is taken as 1988 $\mu$Hz, $T_{\rm sis0}=T_{\rm eS}$. However, $\numax =  1984$ $\mu$Hz makes
$M_{\rm sca}$ and $R_{\rm sca}$  in very good agreement with $M_{\rm sis0}$ and $R_{\rm sis0}$, respectively, and the difference between $T_{\rm sis0}$ and $T_{\rm eS}$ 
is just 21 K.

\subsubsection{KIC 5512589}
We could not determine minima from its observed frequencies. We can only calculate $T_{\rm eVK}$ and $T_{\rm eBV}$ for this star. 
 $T_{\rm eVK}$  is in good agreement with $T_{\rm eS}$.

{ If we take \numax~ as 1277 $\mu$Hz, $M_{\rm sca}$ and $R_{\rm sca}$ are in very good agreement with $M_{\rm lit}$ and $R_{\rm lit}$, respectively.  
This \numax~ also makes $d_{\rm sis}$ equal to $d_{\rm obs}$.}

\subsubsection{KIC 5607242}
KIC 5607242 is an evolved star. It shows mixed modes. 
Its $T_{\rm eS}$ is very close to  $T_{\rm eVK}$. However, $T_{\rm eBV}$ is about 500 K lower than $T_{\rm eVK}$. Furthermore, there is a systematic difference between non-asteroseismic 
and asteroseismic
$T_{\rm eff}$s. The observed value of \numax$ $ (610 $\mu$Hz) gives $T_{\rm sis0}$ as 525 K higher than $T_{\rm eVK}$. If we take \numax$ $ as 640 $\mu$Hz$ $, 
the difference between $T_{\rm sis0}$ and $T_{\rm eVK}$ decreases to 387 K and optimum values are obtained for $M_{\rm sca}$ and $R_{\rm sca}$. 

\subsubsection{KIC 5689820}
KIC 5689820 shows mixed modes.
Observed oscillation frequencies do not allow determination of the frequencies of the minima.
We compute its mass and radius from scaling relations, and photometric effective temperatures from colour.

\subsubsection{KIC 5866724}
The three-planet system KIC 5866724 is an F-like oscillating star (Chaplin et al. 2013). Conventional effective temperatures are very
different and range from 5574 to 6410 K. 
If we reduce \numax$ $ to 1824 $\mu$Hz, $M_{\rm sca}$ is in very good agreement with $M_{\rm sis0}$ and $M_{\rm sis1}$. The same is true for the radii.
The literature values are close to our results.

\subsubsection{KIC 5955122}
KIC 5955122 is an evolved star. The oscillation frequencies for $l=1$ degree show mixed modes. 
KIC 5955122 is also a magnetically active star and has spots (Bonanno et al. 2014). 
It rotates faster than the Sun.
There are two minima (min0 and min1) in its observed oscillation frequencies. 

If we decrease \numax$ $ by an amount of 21 $\mu$Hz, \teff s, masses, and radii computed from different methods are in good agreementbut are larger than the values found in the literature.

\subsubsection{KIC 6106415}
The distance of KIC 6106415 is 41.47 pc. 
Its asteroseismic  \teff s with the observed  value of \numax$ $ (2260 $\mu$Hz) are in good agreement with non-asteroseismic \teff s.
If we take \numax$ $ as 2146 $\mu$Hz, then the agreement between asteroseismic  \teff s and photometric \teff s is particularly excellent.
The modified \numax$ $ also makes $M_{\rm sca}$ and $R_{\rm sca}$ equal to $M_{\rm sis1}$ and $R_{\rm sis1}$, respectively.
Its $d_{\rm sis}$ determined from asteroseismic properties is 39.2 pc.

\subsubsection{KIC 7106245}
We have obtained  \numin$_1$ from its observed oscillation frequencies.
The observed value of \numax$ $ yields a $T_{\rm sis1}$ value 211 K higher than $T_{\rm eS}$. 
If we increase \numax$ $ to 2352 $\mu$Hz, the difference between  $T_{\rm sis1}$  and $T_{\rm eS}$ is 165 K.
Meanwhile,  $M_{\rm sca}$ and $R_{\rm sca}$ become very close to $M_{\rm sis1}$ and $R_{\rm sis1}$, respectively.


\subsubsection{KIC 7341231}
KIC 7341231 is a low-mass red giant and extremely metal-poor star ($[Fe/H]=-1.79$ dex). It may be a halo star (Sharma et al. 2016). 
Deheuvels et al. (2012) obtained the rotation rate of its core from observed oscillation frequencies.
They inferred that the core of KIC 7341231 spins at least five times faster than its surface. 

In this study, we obtain only \numin$_0$ from observed oscillation frequencies. 
If we decrease the observed  \numax$ $ of KIC 7341231 from $\numax=408$ $ $ to $\numax=387$ $\mu$Hz, 
$T_{\rm sis0}$ becomes very close to $T_{\rm eS}$. However, there is a significant difference between $M_{\rm sis0}$ and $M_{\rm sca}$. 
A similar difference is also seen between $R_{\rm sis0}$ and $R_{\rm sca}$. Such differences may arise due to application of 
the relations derived from MS models to the red giants or very low-metallicity of KIC 7341231.

\subsubsection{KIC 7799349}
KIC 7799349 is one of the coolest stars in our sample. 
It is a red giant star and with a spectral effective temperature of 4954 K. 
The \Dnu$ $ versus $\nu$ diagram of KIC 7799349 has interesting features, including mixed modes.
Despite the correction in \numax, there are significant differences between the asteroseismic and non-asteroseismic effective temperature values.
Also, there is no agreement for mass and radius values.
The fitting formulae with \numin$_1$ may not be valid for such cool RG stars. 

\subsubsection{KIC 7871531}
The $T_{\rm eS}$, $T_{\rm eVK}$ and $T_{\rm eBV}$ of KIC 7871531 are slightly different from each other. While $T_{\rm sis0}$ with $\numax=3344$ $\mu$Hz  is very close to $T_{\rm eS}$, $T_{\rm sis1}$  is 133 K greater than $T_{\rm eS}$.
If $\numax$ is taken as $3383$ $\mu$Hz,  $T_{\rm sis0}$ is very close to $T_{\rm eVK}$ and $T_{\rm sis1}$  is nearly the same as $T_{\rm eS}$. 
The modified $\numax$ ($3383$ $\mu$Hz)  makes $R_{\rm sca}$ equal to the other 
values for radius, 0.87. It also yields an $M_{\rm sca}$ in very good agreement with $M_{\rm sis0}$ and  $M_{\rm sis1}$.  

\subsubsection{KIC 7976303}
KIC 7976303 is a sub-giant star. Oscillation frequencies of KIC 7976303 shows mixed modes. 
Slightly modified \numax$ $ (847 $\mu$Hz) makes $M_{\rm sca}$ equal to $M_{\rm sis0}$. $R_{\rm sca}$ from this \numax$ $ is very close to  $R_{\rm lit}$,  $R_{\rm sis0}$, and  $R_{\rm sis1}$.
$d_{\rm sis}$ and  $d_{\rm obs}$ are in very good agreement.

\subsubsection{KIC 8219268}
KIC 8219268 (Kepler-91) is a red giant and a planet host star. 
Kepler-91b has been estimated to be a transiting Jupiter-mass planet (Lillo-Box et al. 2014). 
It is the coolest target star in this study, with a $T_{\rm eS}$  of 4550 K. 
If we slightly decrease the value of \numax$ $, we obtain better agreement between $T_{\rm eS}$ and $T_{\rm sis0}$ and between $R_{\rm sca}$ and $R_{\rm sis0}$. 
However, the difference between $M_{\rm sca}$ and $M_{\rm sis0}$ is high, as with other red giants. 

\subsubsection{KIC 8228742} 
Although the observed value of \numax$ $ (1171  $\mu$Hz) gives $T_{\rm sis0}$ very close to $T_{\rm eS}$, $T_{\rm eVK}$ and $T_{\rm eBV}$, a slightly modified
\numax$ $ (1123  $\mu$Hz) yields $R_{\rm sca}$ exactly equal to $R_{\rm sis0}$ and  $R_{\rm sis1}$.

\subsubsection{KIC 8561221}
KIC 8561221 is a red giant star. Oscillation frequencies of KIC 8561221 with degrees $l=0$, 1, 2 and 3 are observed.

While $T_{\rm eS}$ is in very good agreement with $T_{\rm sis0}$, the difference between $T_{\rm sis0}$ and $T_{\rm sis1}$ is about 
455 K. 
However, the difference between $d_{\rm sis}$ and  $d_{\rm obs}$ is very small.
$M_{\rm sca}$ and $R_{\rm sca}$ are in very good agreement with $M_{\rm lit}$ and $R_{\rm lit}$, respectively.

\subsubsection{KIC 8694723}
The asteroseismic and non-asteroseismic \teff s are in very good agreement. If we slightly modify \numax$ $,
we obtain $M_{\rm sca}$ in perfect agreement with $M_{\rm sis0}$ and  $M_{\rm sis1}$. 

\subsubsection{KIC 8760414}
The $T_{\rm eS}$ of KIC 8760414 is the lowest temperature of five \teff s. If we take the value of \numax$ $ as 2350 $\mu$Hz,
$M_{\rm sca}$ becomes equal to 0.85 \MSbit, the same as $M_{\rm sis0}$ and  $M_{\rm sis1}$.

\subsubsection{KIC 9025370}
The observed value of \numax$ $ (2653 $\mu$Hz) is so low that $T_{\rm sis0}$ is about 534 K higher than $T_{\rm eVK}$ (see Paper III).
If the value of \numax$ $ is taken as 2891 $\mu$Hz, all the asteroseismic masses are equal to 0.97 \MSbit.
In addition, $T_{\rm eS}$ is in very good agreement with $T_{\rm sis0}$ and $T_{\rm sis1}$.
However, there is a significant difference between $d_{\rm sis}$ and  $d_{\rm obs}$.

\subsubsection{KIC 9139163}
KIC 9139163 may be a component of KIC 9139151, another target star in this study (Appourchaux et al. 2015).
If this is true, these stars comprise a rare binary system in which the solar-like oscillating components are observed separately (White et al. 2017). 
Three minima are seen on the \Dnu$ $ versus $\nu$ diagram of KIC 9139163. 
If we take \numax$ $ as 1645 $\mu$Hz, all \teff s are in perfect agreement. This is also the case for the radii.
For the mass, $M_{\rm sca}$ is equal to $M_{\rm sis0}$.

\subsubsection{KIC 9206432}
There are three minima on the $\Dnu$  versus $\nu$ diagram of KIC 9206432. 
$T_{\rm sis0}$, $T_{\rm sis1}$, and $T_{\rm sis2}$ are near to but less than the non-asteroseismic \teff s. The observed value of \numax$ $, 1853 $\mu$Hz,  yields masses and radii very 
different from each other. If we use 1696 $\mu$Hz for \numax, then $T_{\rm sis0}$, $T_{\rm sis1}$, and $T_{\rm sis2}$ are in much better agreement with non-asteroseismic \teff s. 
While $R_{\rm sca}$ is in good agreement with $R_{\rm sis0}$ and $R_{\rm sis1}$, $M_{\rm sca}$ is equal to  $M_{\rm sis1}$.

\subsubsection{KIC 9410862}
\numin$_0$  and \numin$_1$ are from oscillation frequencies with degree $l=1$. 
Using the observed value of \numax$ $ (2261 $\mu$Hz) causes a systematic difference between asteroseismic and non-asteroseismic \teff s.
If we take \numax$ $ as 2175 $\mu$Hz, the systematic differences decrease and excellent agreement is reached for the masses and radii.
While $M_{\rm sca}$, $M_{\rm sis0}$, and  $M_{\rm sis1}$ are very close to 1.01 \MS,  $R_{\rm sca}$, $R_{\rm sis0}$, and  $R_{\rm sis1}$ all become equal to 1.17 \RS.

\subsubsection{KIC 9812850}
$T_{\rm sis1}$  and $T_{\rm sis2}$ are greater than the other \teff s. 
If we take the value of $\numax$ as 1186 $\mu$Hz, $T_{\rm sis1}$  becomes almost equal to $T_{\rm eBV}$, for example. With this modified value, 
$M_{\rm sca}$ is equal to $M_{\rm sis1}$ and $R_{\rm sca}$ is equal to  $R_{\rm sis0}$ and  $R_{\rm sis1}$.

\subsubsection{KIC 10162436} 
Three minima are seen on the $\Dnu$  versus $\nu$ diagram of KIC 10162436. All the asteroseismic \teff s with the observed \numax$ $
are higher than the non-asteroseismic \teff s. If \numax$ $ is taken as 984 $\mu$Hz, $T_{\rm sis0}$, $T_{\rm sis1}$, and $T_{\rm sis2}$ are all very close
to $T_{\rm eBV}$, and $R_{\rm sca}$ is very close to $R_{\rm sis0}$ and  $R_{\rm sis1}$. $M_{\rm sca}$ with the modified \numax$ $ is equal to  $M_{\rm sis1}$. 

\subsubsection{KIC 10454113}
KIC 10454113 is among the stars for which we obtain amazing results.
Its \numax$ $ is determined as 2261 $\mu$Hz. $T_{\rm sis0}$ and $T_{\rm sis1}$  with this value are in good agreement 
with the non-asteroseismic \teff s. If we decrease \numax$ $ to 2175 $\mu$Hz, $M_{\rm sca}$ becomes the same as $M_{\rm sis0}$ and  $M_{\rm sis1}$,
1.16 \MSbit. $R_{\rm sca}$ is in excellent agreement with $R_{\rm sis0}$ and  $R_{\rm sis1}$.

\subsubsection{KIC 11081729}
KIC 11081729 is a very hot solar-like oscillating star.
Although the observed value of \numax$ $ (1990 $\mu$Hz) yields $T_{\rm sis0}$ and $T_{\rm sis1}$  in very good agreement 
with non-asteroseismic \teff s, we modify it (1886 $\mu$Hz) to equalize $M_{\rm sca}$ to  $M_{\rm sis1}$.

\subsubsection{KIC 11244118}
KIC 11244118 is a sub-giant star which is observed by {\it Kepler} with short cadence (58.85 s, Gilliland et al. 2013). 
Karoff et al. (2013)  proposed that KIC 11244118 is an active star. 
Observed \numax$ $ is 1420 $\mu$Hz. $T_{\rm sis0}$ is in good agreement with $T_{\rm eS}$ and $T_{\rm eVK}$.
For the good agreement between asteroseismic and non-asteroseismic masses and radii, the required value for \numax$ $ is 1377 $\mu$Hz.

\subsubsection{KIC 11295426}
KIC 11295426 (Kepler-68) is a planet host star. It has three planets (Gilliand et al. 2013). 
If  \numax$ $ is taken as 2085 $\mu$Hz, asteroseismic effective temperatures (except $T_{\rm sis1}$) are in excellent agreement with non-asteroseismic effective temperatures. 
Asteroseismic masses are in agreement with themselves and with the literature value.

\subsubsection{KIC 11414712}
KIC 11414712 is in sub-giant evolution stage with mixed modes.
Its \numax$ $ is taken as 702 $\mu$Hz. Two minima are obtained from observed frequencies. 
$T_{\rm sis0}$ is slightly greater than non-asteroseismic \teff s but $M_{\rm sca}$ is very close to $M_{\rm sis0}$ and equal to  $M_{\rm sis1}$.

\subsubsection{KIC 11772920}
Unfortunately, only \numin$_0$ is available from the $\Dnu$  versus $\nu$ diagram of KIC 11772920. 
Its observed \numax$ $ (3439 $\mu$Hz) 
gives very high $T_{\rm sis0}$. The difference between $T_{\rm sis0}$ and $T_{\rm eS}$ is about 581 K. If \numax$ $ is taken as  3643 $\mu$Hz, this difference 
is reduced to 192 K. Furthermore, $M_{\rm sca}$ with this value of  \numax$ $ becomes very close to $M_{\rm sis0}$ and  $R_{\rm sca}$ is equal to $R_{\rm sis0}$.

\subsubsection{KIC 11807274}
KIC 11807274 is an evolved star and hosts two planets (Chaplin et al. 2013). 
The observed value of \numax$ $ gives $T_{\rm sis1}$  in very good agreement with $T_{\rm eS}$. However, $T_{\rm sis0}$ is 180 K is less than $T_{\rm eS}$.
In order to obtain agreement between $R_{\rm sca}$, $R_{\rm sis0}$, and  $R_{\rm sis1}$, \numax$ $ is decreased to 1445 $\mu$Hz. With this value of  \numax$ $,
$M_{\rm sca}$ is equal to  $M_{\rm sis1}$.  

\subsubsection{KIC 12069424}
One of the brightest $Kepler$ targets is KIC 12069424 (16 Cyg A), an evolved star. 
Metcalfe et al. (2012) identify 46 oscillation frequencies, including the modes with $l=3$.
They determine the fundamental properties of this star by constructing interior models with different stellar evolution codes. 
If we slightly increase (31 $\mu$Hz) the value of \numax, $R_{\rm sca}$ becomes equal to $R_{\rm sis0}$ and  $R_{\rm sis1}$.
With the modified \numax$ $, $M_{\rm sca}$ is equal to $M_{\rm sis0}$ and  $M_{\rm sis1}$.
The asteroseismic distance of KIC 12069424 is almost the same as the observed distance. 

\subsubsection{KIC 12069449}
KIC 12069449 (16 Cyg B) is an MS star and hosts a planet. 
Its observed \numax$ $ is 2552  $\mu$Hz. This value yields a $T_{\rm sis0}$ lower than the non-asteroseismic \teff s and a $T_{\rm sis1}$  slightly higher than them.
If we take \numax$ $ as 2485 $\mu$Hz, $T_{\rm sis0}$ is very close to $T_{\rm eBV}$ and $M_{\rm sca}$ is very close to $M_{\rm sis0}$ and $M_{\rm sis1}$.
With the modified \numax$ $, the asteroseismic radii are in very good agreement.

\subsubsection{HD 2151}
HD 2151 ($\beta$ Hyi, HR 98, HIP 2021) is a bright sub-giant star. 
Bedding et al. (2007) have observed frequencies of this star using high-precision velocity observation with HARPS and UCLES spectrographs. 
They also identified 28 modes with degrees $l=0$, 1 and 2. 
$M_{\rm sca}$ is in very good agreement with $M_{\rm sis0}$ and  $M_{\rm sis1}$.
This star is very close at a distance of 7.46 pc (from {\it Hipparcos} parallax). We determined its asteroseismic distance to be 7.08 pc, very similar to the observed value. 

\subsubsection{HD 43587}
The evolved star HD 43587  shows mixed modes in its oscillation frequencies, which were observed by {\it CoRoT} (Boumier et al. 2014). 
If we decrease observed \numax$ $ to 2215 $\mu$Hz, asteroseismic and non-asteroseismic \teff s 
are in very good agreement.
The asteroseismic radii of HD 43587 are in excellent agreement and $M_{\rm sca}$ is very close to $M_{\rm sis0}$ and  $M_{\rm sis1}$.

\subsubsection{HD 146233} 
Target star HD 146233 (18 Sco) was identified as a solar twin (Porto de Mello \& de Silva 1997). 
The oscillation frequencies of this star are obtained by HARPS spectrometer. 
\teff s are very different from each other, but $T_{\rm eS}$ is very close to $T_{\rm eBV}$. 
If we take $\numax$ as $2970$ $\mu$Hz, $T_{\rm sis1}$ is in good agreement with $T_{\rm eVK}$.
This value of \numax$ $ yields a $M_{\rm sca}$ in better agreement with $M_{\rm sis0}$ and  $M_{\rm sis1}$. All the asteroseismic radii are 
in excellent agreement with 
the interferometric radius obtained by Bazot et al. (2011) as 
1.010$\pm$ 0.009 \RSbit.

\subsubsection{Procyon A}
Procyon is a spectroscopic binary system and Procyon A shows solar-like oscillations. 
The oscillation frequencies of Procyon A are obtained through a multisite spectroscopic campaign (Bedding et al. 2010). 
It is the closest star (3.51 pc, from {\it Hipparcos} parallax) in our sample. 
Its mass and radius are determined with non-asteroseismic methods using
{\it Hubble Space Telescope} and ground-based astrometric data.  \\
The observed mass of sub-giant Procyon A is 1.478 $\pm$ 0.012 \MS (Bond et al. 2015). 
The interferometric radius of Procyon A is 2.031 $\pm$ 0.013 \RS (Aufdenberg, Ludwig \& Kervella 2005).
In this study, the observed \numax$ $ of Procyon A is taken as 983 $\mu$Hz. 
When we use the modified \numax$ $ in scaling relations, asteroseismic mass and radius are in excellent agreement with non-asteroseismic values.
Also, asteroseismic effective temperatures are compatible with $T_{\rm eS}$.  
This value of  \numax$ $ is very close to mean of the results from photometric and spectroscopic observations.


{
In Fig. 8, $\Delta \nu$ versus $\nu$ graph is plotted for the observed oscillation frequencies of Procyon A with $l=0$
(Bedding et al. 2010).
The minima at $\numin_1=1126.7$ $\mu$Hz in Fig. 8 is min1. $T_{\rm sis1}$ from $\nu_{\rm min1}$ is 6597 K and very close to 
$T_{\rm eBV}$, $T_{\rm eVK}$, and $T_{\rm eS}$. $M_{\rm sis1}$ and $R_{\rm sis1}$ computed from $\nu_{\rm min1}$ using equations (1) and (4) are also in very good agreement
with the mass and radius derived from non-asteroseismic analysis, respectively. There are two more minima in the low frequency domain of 
Fig. 8. Their frequencies are 849.1 and 739.2 $\mu$Hz. If we compare $T_{\rm sis2}$ computed from the frequencies of the minima, we see that 
they give $T_{\rm sis2}$ as 6725 and 6550 K, respectively. This implies that $T_{\rm sis2}$ from $\nu_{\rm min2}=739.2$ $\mu$Hz is in better agreement with \teff$ $ from
the other methods than $T_{\rm sis2}$ from 849.1 $\mu$Hz. However, Procyon A is among the hottest stars
in our sample and equation (A4) is questionable for such stars. Procyon A deserves much more detailed analysis than the present analysis.
}
 
\begin{figure}
\includegraphics[width=101mm,angle=0]{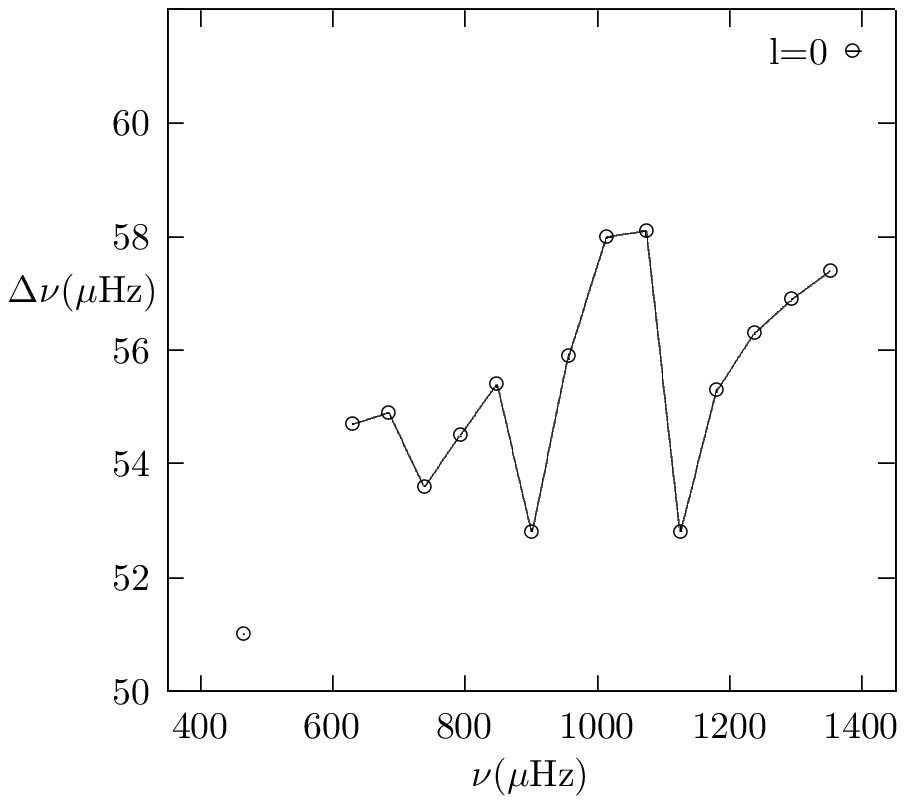}
\caption{$\Delta \nu$ is plotted with respect to $\nu$ for the observed oscillation frequencies of Procyon A for $l=0$
(Bedding et al. 2010). 
Three minima appear for $l=0$; one is about $\nu=1126.7$ $\mu$Hz and the others are about 849.1 and 739.2 $\mu$Hz. 
}
\end{figure}

\section{Consequences of the findings }
\subsection{Chemical evolution of Galactic Disc}
Metallicity is very important for our understanding of stellar evolution in two respects. First
it dramatically influences the inner core as much as the outer regions. Secondly, it in turn implicitly 
represents stellar age because heavy element abundances in Galactic Disc increase in time. However, to date, the relation between heavy metal abundances and time has not been fully elucidated. 

In Fig. 9, $\log Z_{\rm o}$ is plotted against $\log (t_9)$ and yields very striking results. There is a continuous metal enrichment
in all stages for both MS and post-MS stars. Different ages have different metallicity intervals. There are very 
old stars with high metallicity. For example, some of the stars with ages of about 7 Gyr have $Z_{\rm o}$ ($\log Z_{\rm o}=-1.6$)
of about 0.022. The maximum metallicity of the youngest stars (about 0.025) is slightly higher than this value.  
This implies that the maximum metallicity does not change very rapidly over time, at least during the last 7 Gyr. However,
for all ages, there is a minimum metallicity and it changes so rapidly that there is no young star with low metallicity in our sample.
These confirmations may be very important for our understanding of the chemical evolution of the Galaxy as 
a result of repeatedly reprocessing stellar material.  
\begin{figure}
\includegraphics[width=101mm,angle=0]{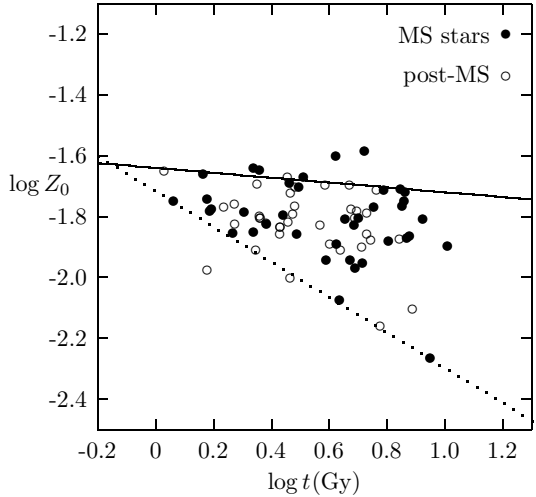}      
\caption{$\log Z_{0}$ is plotted \wrt logarithm of age.
The filled circles are for MS and the circles are for post-MS stars.
The solid line represents time variation of maximum metallicity ($-0.08\log t_9 -1.64$).
The dotted line shows time variation of minimum metallicity ($-0.58\log t_9 -1.72$).
It is very interesting that the relation between the logarithm of the lowest metallicity and $\log t_9$ behaves
as $Z_{\rm 0min}\propto t^{-0.58}$.
}
\end{figure}
\subsection{Gyrochronology}
The magnetic braking mechanism is responsible for the slow rotation of late-type stars with 
convective envelope (Epstein \& Pinsonneualt 2014).
For most of the target stars, rotational velocity ($v\sin i$) is available in the literature.
The  $v\sin i$ values are taken from Pinsonneault et al. (2014).
Five stars with ($\teff> 6165$ K) are fast rotators, $18 <v\sin i<25$ km\,s$^{-1}$.
These stars must have a thin convective zone during their MS phase (van Saders et al. 2016).
$v\sin i$ shows mass/colour dependence (Barnes 2007; Mamajek \& Hillenbrand 2008; Angus et al. 2015).
In order to obtain the time dependence of $v\sin i$, the mass dependence must first be subtracted.
We confirm that the mass dependence can be represented by a linear function of $\log M$ given as
\begin{equation} 
f_M = (3.29\pm0.46)\log(M/\MS)+(0.343\pm0.048) 
\end{equation}
from a graph of $\log (v\sin i)$ versus $\log (M_{\rm sca})$.
In Fig. 10, $\log (v\sin i)-f_M$ is plotted \wrt $\log (t_9)$.
The fitted line is
\begin{equation} 
g_t= (-0.51\pm0.13)\log(t_9)+(0.31\pm0.08)
\end{equation}
where $g_t= \log (v\sin i)-f_M$.
This implies that $v\sin i\propto t^{-0.51\pm0.13}$. This result is in very good agreement with
the famous Skumanich relation (Skumanich 1972). In Fig. 10, a very striking result emerges when
we compare MS and post-MS stars. Particularly at ages greater than 2.5 Gyr,
the MS stars are rotating faster than post-MS stars with the same mass. The mean difference between
$\log (v\sin i) - f_M$ of the MS stars (0.132) and that of the post-MS (-0.143) stars is about 27 
per cent. This implies that
MS stars rotate about 2.3 km\,s$^{-1}$ faster than post-MS stars with the same mass.

A very similar result is found for the relation between age and rotation if we first subtract the 
$B-V$ colour dependence (Mamajek \& Hillenbrand 2008) of $ \log (v\sin i)$  ($f_{BV}$) and then
utilize time dependence of $\log (v\sin i/($km\,s$^{-1}))-f_{BV}$. This approach yields the time dependence 
of projected rotation speed as $v\sin i\propto t^{-0.58\pm0.14}$.




\begin{figure}
\includegraphics[width=101mm,angle=0]{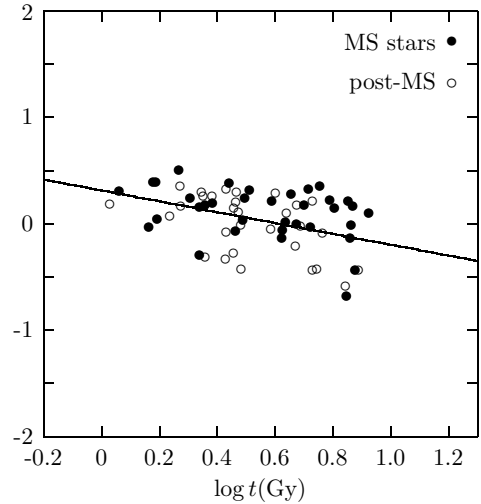}
\caption{$\log (v\sin {i})-f_M$ is plotted \wrt logarithm of age.
The filled circles are for MS and the circles are for post-MS stars.
The MS stars rotate slightly faster than the post-MS stars with the same mass.
The solid line shows mean time variation of $\log (v\sin {i})-f_M$: $\log (v\sin {i})-f_M=-0.51\log t_9+0.31$.
}
\end{figure}

\subsection{Comparison of the observed and asteroseismic parallaxes}
{
The zero-point offset between the observed and asteroseismic parallaxes is considered in many papers in the literature 
(Sahlholdt et al. 2018; Zinn et al. 2018; Khan et al. 2019; Hall et al. 2019).
Most of these studies find the observed parallax less than the asteroseismic parallax: $\Delta \pi=\pi_{\rm obs}-\pi_{\rm sis}$, about -0.05 mas. 
If we use observed \numax~ in scaling relations,
we find that  $\Delta \pi=-0.2835$ mas. If we use modified \numax, then $\Delta \pi=-0.439 $ mas.
This a very interesting result. While we obtain very good agreement between $M$, $R$, and $\teff$ computed from different methods using modified \numax,
the discrepancy between $\pi_{\rm obs}$ and $\pi_{\rm sis}$ increases.
It is quite difficult to find the real reason behind the zero-point offset.
If we use the pure asteroseismic parameters obtained using \numin$_1$, including $T_{\rm sis1}$, for example,
$\Delta \pi$ reduces to 0.005 mas.

\begin{figure}
\includegraphics[width=101mm,angle=0]{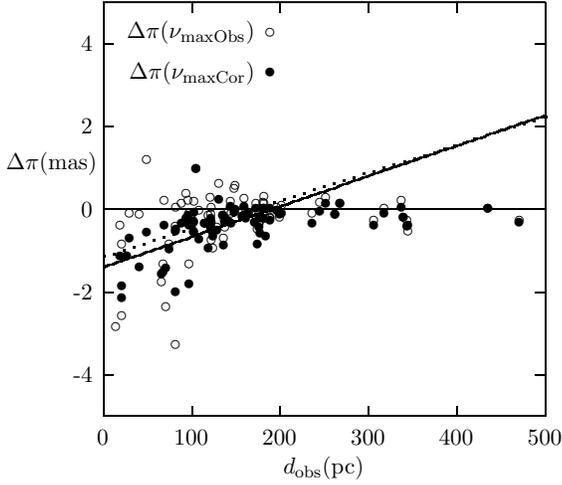}
\caption{$\Delta \pi$ is plotted \wrt $d_{\rm obs}$.
The circles and filled circles are for the asteroseismic parallax from the observed and corrected $\nu_{\rm max}$, respectively.
The dotted and thick solid lines are the fitted lines for these data with $d_{\rm obs}<200$ pc.
}
\end{figure}
Most of the studies on the comparison of {\it Gaia} parallaxes with parallaxes derived from asteroseismic or non- asteroseismic radius estimation 
tries to obtain a constant difference between them. 
The difference between the parallaxes corresponding to the  $d_{\rm obs}$ and  $d_{\rm sis}$
given in Table B1 is plotted as a function of  $d_{\rm obs}$ in Fig. 11. There appears a definite linear relationship  between $\Delta \pi$ and $d_{\rm obs}$ for 
$d_{\rm obs} < 200$ pc. A very similar relationship is obtained also for the asteroseismic parallax derived from the observed $\nu_{\rm max}$.
We also notice that in both case $\Delta \pi$ is very close to the horizontal line for $\Delta \pi=0$ for the range $d_{\rm obs} > 200$ pc.
Sahlholdt et al. (2018) finds similar dependence for $\Delta \pi$ with the distance. %
}


{The greatest difference between the parallaxes appears for the stars with \numax$>$2000 $\mu$Hz.   
9 of the 10 stars with $|\Delta \pi|>1$ have \numax$>$2000 $\mu$Hz. The mean difference ($\Delta \pi$) is about -0.06 mas for the stars with \numax$<$2000 $\mu$Hz.
There might be several sources of discrepancy between parallaxes other than \numax$ $ and $\pi_{\rm G}$:\\
i) asteroseismic scaling relation for $R$,\\
ii) $V$ and/or interstellar extinction,\\
iii) $BC$,\\
iv)  the solar values of \numax$ $ and \Dnu.\\
In order to remove the difference between $\pi_{\rm G}$ and  $\pi_{\rm sis}$, the required value of $\nu_{\rm max{\sun}}$, for e.g. 
is about $3000$ $\mu$Hz rather than $3050$ $\mu$Hz we adopted.
}
  
\section{Conclusions}
The He {\scriptsize II} ionization zone causes \Dnu$ $ to have an oscillatory component in the spacing of oscillation frequencies.  
Both models and observations have several minima in the \Dnu$ $ versus $\nu$ graph.
In our previous papers, we have already shown that the frequencies of these minima have very strong diagnostic potential
for determination of the fundamental properties of solar-like oscillating stars. The \teff s$ $  of about 90 stars were obtained from their
oscillation frequencies in Paper III. In this study, we develop new methods for determining $M$, $R$, $g$, and age using 
models constructed with the {\small MESA} code and apply the methods to these stars.

The precision of new relations for $M$ and $R$ are comparable to the new scaling relations with $\Gamma_{\rm \negthinspace 1s}$ presented in Paper III. However,
uncertainties in $M_{\rm sca}$ and $R_{\rm sca}$ are not low enough for many stars because $\numax$ cannot be obtained 
very accurately for many of the solar-like oscillating stars. Furthermore, 
$M_{\rm sca}$ is a very strong function of $\numax$.
Therefore, $M_{\rm sis0}$ and  $M_{\rm sis1}$ are much more accurate than
$M_{\rm sca}$. 
The mass and radius range of the targets are determined to be [0.8,1.8] \MS and [0.85,13] \RSbit, respectively.
The maximum difference between
$M_{\rm sis1}$ and $M_{\rm sis0}$ is about 4 per cent and 
very precise results are obtained for 38 of these stars:
$\mid M_{\rm sis0}- M_{\rm sis1}\mid <0.024$ \MS and $\mid R_{\rm sis0}- R_{\rm sis1}\mid< 0.007$ \RSbit.
This shows that frequencies \numin$_0$, \numin$_1$, and \numin$_2$ have strong diagnostic potentials for 
the determination of fundamental properties of stars, including mass and radius. 

We also compute the age and initial metallicity of the target stars and obtain very interesting results regarding the relation between age and metallicity (see Fig. 9). It seems that every time has its own $Z$ interval in Galactic Disc and there is a very clear relation
in particular between the minimum $Z$ and age.

Once we obtain $M_{\rm sis0}$ and $R_{\rm sis0}$, 
we modify the value of \numax$ $ so that all three quantities
\teff, $M_{\rm sca}$, and $R_{\rm sca}$ are in better agreement with $T_{\rm sis0}$, $M_{\rm sis0}$, and $R_{\rm sis0}$, respectively.
The same is true for $M_{\rm sis1}$ and $R_{\rm sis1}$, of course. 
In that way, we can precisely determine the value of \numax.

We also obtain an inverse relation between age and $v \sin(i)$: $v \sin(i)\propto t^{-0.51}$. This relation is almost the
same as the well-known Skumanich relation derived for the relatively young low-mass stars.

{
We also compare the asteroseismic and {\it Gaia} parallaxes and find an offset. The offset seems to be distance dependent.
Further and more detailed analysis is needed to find the real reason behind the offset. 
}

\section*{Acknowledgements}
We acknowledged Dr. J. L. van Saders for her discussions on gyrochronology of late-type stars.

\appendix

\section{Corrections in scaling relations} 
We compute \numax$ $  of the {\small MESA}  models by assuming $f_\nu=1$:
\begin{equation}
\frac{\nu_{\rm max}}{\nu_{\rm max\sun}}= 
\frac{g/g_{\sun}}{(\teff/\teff_{\sun})^{0.5}} 
\left(\frac{\Gamma_{\rm \negthinspace 1s}}{\Gamma_{\rm \negthinspace 1s\sun}}\right)^{0.5}.
\end{equation}
The relation between \teff$ $ and $\Delta n_{\rm xi}$ is plotted in Fig. A1. 
The fitting curves for $T_{\rm sis0}$, $T_{\rm sis1}$ and $T_{\rm sis2}$ are given below:
\begin{figure}
\includegraphics[width=101mm,angle=0]{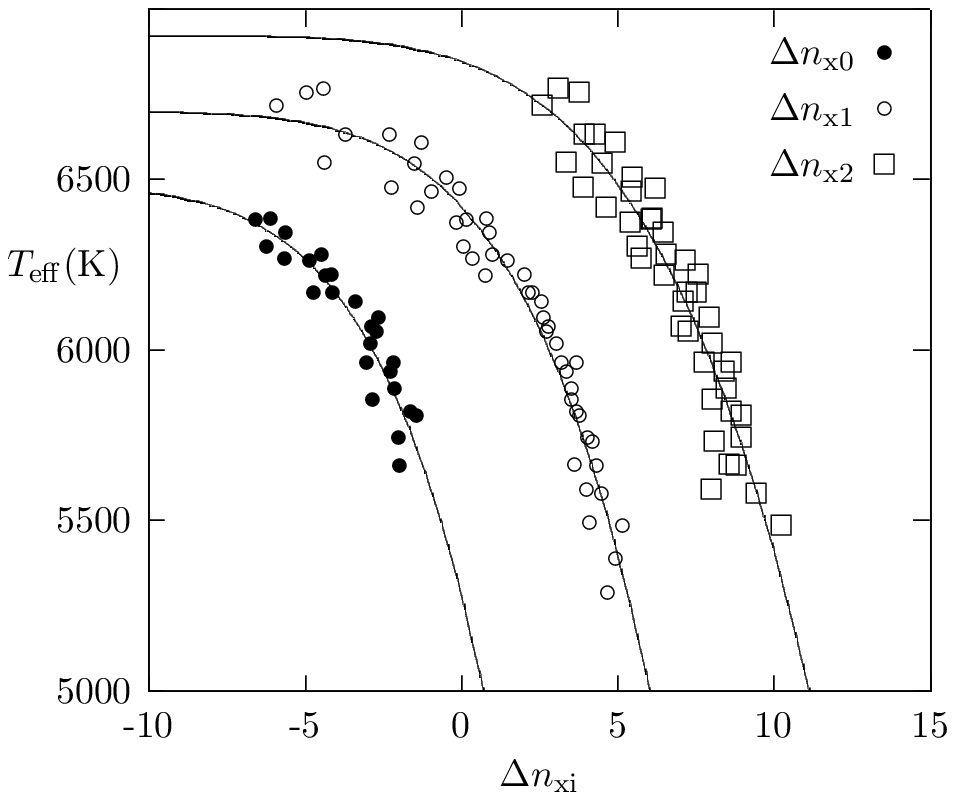}
\caption{Effective temperature is plotted \wrt $\Delta n_{\rm x0}$ (filled circles), $\Delta n_{\rm x1}$ (circles), and $\Delta n_{\rm x2}$ (squares).
}
\end{figure}

\begin{equation}
\frac{T_{\rm sis0}(\Delta n_{\rm x0})}{\rm T_{\rm eff \sun}}=1.121-3.235\times 10^{-9}(\Delta n_{\rm x0}+20)^{6},
\end{equation}
\begin{equation}
\frac{T_{\rm sis1}(\Delta n_{\rm x1})}{\rm T_{\rm eff \sun}}=1.159-2.530\times 10^{-8}(\Delta n_{\rm x1}+15)^{5.34}
\end{equation}
and
\begin{equation}
\frac{T_{\rm sis2}(\Delta n_{\rm x2})}{\rm T_{\rm eff \sun}}=1.198-5.896\times 10^{-7}(\Delta n_{\rm x2}+10)^{4.34}.
\end{equation}

\newpage
\onecolumn
\section{Basic asteroseismic and non-asteroseismic properties of {\it K\lowercase{epler}} and {\it C\lowercase{o}R\lowercase{o}T} Targets}

\small\addtolength{\tabcolsep}{-4pt}

\begin{landscape}

\begin{longtable}{@{}rrrrrrrrrrrrrrrrrrrrrrrrrrrrrrr@{}}
\caption[Continued.]{ 
Basic properties of {\it Kepler} and {\it CoRoT} targets. Columns are organized as star name, frequency of maximum amplitude, 
reference frequencies for min0, min1, and min2, mean large and small separations between oscillation frequencies, effective 
temperatures [from spectra, $V-K$ and $B-V$ colours (see section 2 in Paper III) and from minima min0, min1, and min2 (see section 
3 in Paper III), respectively], masses and radii (from new scaling relation (see section 3 in Paper III), from min0 and min1 
and from literature, respectively), distances (from literature and from seismic), surface gravities [from new scaling relations 
(see section 3 in Paper III), from seismic (see Section 2) and from spectra, respectively], ages [from seismic (see Section 2) 
and from method that based on mass, radius and metallicity (see \yildiz et al. 2014b for details), metallicities (from spectra and 
from computed with diffusion (see Section 3)] and numbers of references. Second row describes uncertainities of basic properties. 
Unlike others, for Procyon A, mass and radius are the observed values, not the model values in the literature. Sun is denoted by 
${\bigodot}$ symbol at the end of the table.
}\\
\hline
Star    & \numax & $\nu_{\rm min0}$ & $\nu_{\rm min1}$ & $\nu_{\rm min2}$ &  \Dnu & $\braket{\delta \nu_{02}}$ & $T_{\rm eS}$ &  $T_{\rm eVK}$ &  $T_{\rm eBV}$ & $T_{\rm sis0}$ & $T_{\rm sis1}$ & $T_{\rm sis2}$ & $M_{\rm sca}$ &$ M_{\rm sis0}$ &$ M_{\rm sis1}$ & $M_{\rm lit}$ & $R_{\rm sca}$ &$ R_{\rm sis0}$ &$ R_{\rm sis1}$ & $R_{\rm lit}$  & $d_{\rm obs}$ & $d_{\rm sis}$& $g_{\rm sca}$ &$ g_{\rm sis}$ & $g_{\rm spc}$ & $t_{\rm sis}$ & $t_{\rm yil}$ & $Z_{\rm s}$ & $Z_{\rm o}$ & Ref \\
		& ($\mu$Hz) & ($\mu$Hz)   & ($\mu$Hz)  & ($\mu$Hz)  &  ($\mu$Hz) & ($\mu$Hz) & (K) &  (K)   &  (K)   & (K)   &  (K)   & (K)   & (\MS)     &  (\MS)  &  (\MS)     &  (\MS)   & (\RS)     &  (\RS)  &  (\RS)     &  (\RS)   & (pc) & (pc) & & & & (Gyr)&  (Gyr) &     &     &   \\
\hline
\endfirsthead

\caption[-- continued]{-- continued}\\
\hline
Star    & \numax & $\nu_{\rm min0}$ & $\nu_{\rm min1}$ & $\nu_{\rm min2}$ &  \Dnu & $\braket{\delta \nu_{02}}$ & $T_{\rm eS}$ &  $T_{\rm eVK}$ &  $T_{\rm eBV}$ & $T_{\rm sis0}$ & $T_{\rm sis1}$ & $T_{\rm sis2}$ & $M_{\rm sca}$ &$ M_{\rm sis0}$ &$ M_{\rm sis1}$ & $M_{\rm lit}$ & $R_{\rm sca}$ &$ R_{\rm sis0}$ &$ R_{\rm sis1}$ & $R_{\rm lit}$  & $d_{\rm obs}$ & $d_{\rm sis}$& $g_{\rm sca}$ &$ g_{\rm sis}$ & $g_{\rm spc}$ & $t_{\rm sis}$ & $t_{\rm yil}$ & $Z_{\rm s}$ & $Z_{\rm o}$ & Ref \\
		& ($\mu$Hz) & ($\mu$Hz)   & ($\mu$Hz)  & ($\mu$Hz)  &  ($\mu$Hz) & ($\mu$Hz) & (K) &  (K)   &  (K)   & (K)   &  (K)   & (K)   & (\MS)     &  (\MS)  &  (\MS)     &  (\MS)   & (\RS)     &  (\RS)  &  (\RS)     &  (\RS)   & (pc) & (pc) & & & & (Gyr)&  (Gyr) &     &     &   \\
\hline
\endhead

\hline
\endfoot

{ 1435467} &   1335.0 &   1626.4 &   1274.0 &      --- &   70.9 &    4.8 &   6264 &   6224 &   6587 &   6178 &   6322 &    --- &     1.27 &     1.24 &     1.27 &     1.27 &     1.66 &     1.65 &     1.66 &     1.64 &   137.22 &   133.19 &     4.10 &     4.10 &     4.09 &   2.92 &   2.92 &   0.0118 &   0.0188 & 2,13,17 \\
              &     40.0 &     16.3 &     12.7 &      --- &    0.8 &    --- &     60 &     51 &    185 &     87 &     92 &    --- &     0.19 &     0.02 &     0.02 &     0.05 &     0.10 &     0.01 &     0.01 &     0.03 &     0.57 &     1.22 &     0.01 &     0.01 &     0.03 &   0.21 &    --- &   0.0009 &   0.0009 &   39    \\
{ 2837475} &   1500.0 &   2265.6 &   1689.9 &   1276.8 &   75.2 &    6.7 &   6462 &   6545 &   6488 &   6462 &   6593 &   6689 &     1.49 &     1.54 &     1.49 &     1.39 &     1.67 &     1.66 &     1.65 &     1.59 &   121.24 &   115.34 &     4.17 &     4.18 &     3.95 &   1.15 &   1.02 &   0.0111 &   0.0178 & 2,17,39 \\
              &     54.0 &     22.7 &     16.9 &     12.8 &    1.3 &    --- &    125 &     54 &    144 &     10 &     41 &     61 &     0.31 &     0.03 &     0.03 &     0.06 &     0.13 &     0.02 &     0.02 &     0.03 &     0.43 &     5.85 &     0.02 &     0.02 &     0.23 &   0.11 &   0.19 &   0.0031 &   0.0037 &   40    \\
{ 3424541} &    722.0 &   1046.6 &    755.4 &      --- &   41.1 &    4.7 &   6165 &   6249 &   6322 &   6420 &   6491 &    --- &     1.74 &     1.80 &     1.74 &     1.64 &     2.65 &     2.65 &     2.64 &     2.53 &   317.59 &   307.40 &     3.83 &     3.84 &     3.90 &   1.07 &   1.84 &   0.0139 &   0.0222 & 2,17,39 \\
              &     55.0 &     10.5 &      7.6 &      --- &    1.1 &    --- &    108 &     73 &    249 &     44 &    116 &    --- &     0.63 &     0.04 &     0.04 &     0.04 &     0.37 &     0.03 &     0.03 &     0.07 &     2.83 &    10.74 &     0.02 &     0.02 &     0.21 &   0.15 &   0.52 &   0.0038 &   0.0263 &   40    \\
\hline
\end{longtable}
~~\\
{\it Note}: Ref.  $-$ .... 54: Brewer et al. (2016), 55: Frasca et al. (2016).
See Paper III for the other references.
\end{landscape}


\begin{thebibliography}{99}
\bibitem[\protect\citeauthoryear{Angus et al.}{2015}]{angus} Angus R., Aigrain S., Foreman-Mackey D., McQuillan A., 2015, MNRAS, 450, 1787
\bibitem[\protect\citeauthoryear{Appourchaux et al.}{2015}]{Appour} Appourchaux T. et al., 2015, A\&A, 582, A25
\bibitem[\protect\citeauthoryear{Arentoft et al.}{2008}]{Arentoft} Arentoft T. et al., 2008, ApJ, 687, 1180
\bibitem[\protect\citeauthoryear{Aufdenberg,  Ludwing \& Kervella}{2015}]{aufden} Aufdenberg J. P., Ludwing H. G.,  Kervella P.,  2005, ApJ, 633, 424
\bibitem[\protect\citeauthoryear{Bahcall, Serenelli \& Pinsounneault}{2004}]{bah04} Bahcall J. N., Serenelli A. M., Pinsonneault M., 2004, ApJ, 614, 464
\bibitem[\protect\citeauthoryear{Ballard et al.}{2014}]{bal} Ballard S. et al., 2014, ApJ, 790, 12
\bibitem[\protect\citeauthoryear{Barnes}{2007}]{barnes} Barnes S. A., 2007, ApJ, 669, 1167
\bibitem[\protect\citeauthoryear{Bazot et al.}{2011}]{baz11} Bazot M. et al., 2011, A\&A, 526, L4
\bibitem[\protect\citeauthoryear{Bedding et al.}{2007}]{Bedding} Bedding T. R. et al., 2007, ApJ, 663, 1315
\bibitem[\protect\citeauthoryear{Bedding et al.}{2010}]{bedding} Bedding T. R. et al., 2010, ApJ, 713, 935
\bibitem[\protect\citeauthoryear{Benomar et al.}{2014}]{ben} Benomar O., Masuda K., Shibahashi H., Suto Y., 2014, PASJ, 66, 94
\bibitem[\protect\citeauthoryear{Bonanno et al.}{2014}]{bonanno} Bonanno A., Frohlich H. E., Karoff C., Lund M. N., Corsaro E. Frasca A., 2014, A\&A, 569, A113
\bibitem[\protect\citeauthoryear{Bond et al.}{2015}]{bond} Bond H. et al, 2015, ApJ, 813, 106
\bibitem[\protect\citeauthoryear{Boumier et al.}{2014}]{boumier} Boumier P. et al., 2014, A\&A, 564, A34
\bibitem[\protect\citeauthoryear{Brewer et al.}{2016}] {brew} Brewer, J. M., Fischer, D. A., Valenti, J. A., Piskunov, N., 2016, ApJS, 225, 32
\bibitem[\protect\citeauthoryear{Bruntt et al.}{2012}]{bru12} Bruntt H. et al., 2012, MNRAS, 423, 122
\bibitem[\protect\citeauthoryear{Chaplin et al.}{2013}]{chaplin} Chaplin W. J. et al., 2013, ApJ, 766, 101
\bibitem[\protect\citeauthoryear{Deheuvels et al.}{2012}]{dehe12} Deheuvels S. et al., 2012, ApJ, 756, 19
\bibitem[\protect\citeauthoryear{Edvardsson et al.}{1993}]{edv} Edvardsson B., Andersen J., Gustafsson B., Lambert D. L., Nissen P. E., Tomkin J., 1993, A\&A, 275, 101 
\bibitem[\protect\citeauthoryear{Epstein \& Pinsonneault}{2014}]{epstein} Epstein C. R., Pinsonneault M. H., 2014, ApJ, 780, 159
\bibitem[\protect\citeauthoryear{Frasca et al.}{2016}]{Frasca et al.} Frasca A. et al., 2016, A\&A, 594, A39
\bibitem[\protect\citeauthoryear{Gaia Coll}{2018}]{Gaia Collaboration} Gaia Collaboration 2018, A\&A, 616, A1
\bibitem[\protect\citeauthoryear{Gilliland et al.}{2013}]{gil13} Gilliland R. L. et al., 2013, ApJ, 766, 40
\bibitem[\protect\citeauthoryear{Gruyters, Nordlander \& Korn}{2014}]{gruy14} Gruyters P., Nordlander T.,  Korn A. J., 2014, A\&A, 567, A72
\bibitem[\protect\citeauthoryear{Gruyters et al.}{2016}]{gruy16} Gruyters P. et al., 2016, A\&A, 589, A61
\bibitem[\protect\citeauthoryear{Hall et al.}{2019}]{Hall} Hall, O. J., Davies, G. R., Elsworth, Y. P., et al. 2019, MNRAS, 486, 3569
\bibitem[\protect\citeauthoryear{Howell et al.}{2012}]{how12} Howell R. L. et al., 2012, ApJ, 746, 123
\bibitem[\protect\citeauthoryear{Huber et al.}{2011}]{Huber} Huber D. et al., 2011, ApJ, 743, 143
\bibitem[\protect\citeauthoryear{Kallinger et al.}{2010}]{kal10} Kallinger T. et al., 2010, A\&A, 522, A1
\bibitem[\protect\citeauthoryear{Karoff et al.}{2013}]{kar13} Karoff C. et al., 2013, MNRAS, 433, 3227
\bibitem[\protect\citeauthoryear{Khan et al.}{2019}]{Khan} Khan S. et al., 2019, A\&A, 628, A35
\bibitem[\protect\citeauthoryear{King et al.}{1998}]{king98} King J. R., Stephens A., Boesgaard A. M., Deliyannis C., 1998, ApJ, 115, 666
\bibitem[\protect\citeauthoryear{Korn et al.}{2007}]{korn}Korn A. J., Grundahl F., Richard O., Mashonkina L., Barklem P. S., Collet R., Gustafsson B., Piskunov N., 2007, ApJ, 671, 402
\bibitem[\protect\citeauthoryear{Lillo-Box et al.}{2014}]{lil} Lillo-Box J. et al., 2014, A\&A, 562, A109
\bibitem[\protect\citeauthoryear{Mamajek \& Hillenbrand}{2008}]{mamajek} Mamajek E. E., Hillenbrand L. A., 2008, ApJ, 687, 1264
\bibitem[\protect\citeauthoryear{Mathur et al.}{2012}]{Mathu} Mathur S. et al., 2012, ApJ, 749, 152
\bibitem[\protect\citeauthoryear{Metcalfe et al.}{2012}]{Metcalfe} Metcalfe T. S. et al., 2012, ApJ, 748, L10
\bibitem[\protect\citeauthoryear{Michaud et al.}{2012}]{Michaud} Michaud G., Proffitt C. R., 1993, in Weiss W. W., Baglin A., eds, ASP Conf. Ser. Vol. 40, IAU Colloq.
137: Inside the Stars. Astron. Soc. Pac., San Francisco, p. 246
\bibitem[\protect\citeauthoryear{Molenda-{\.{Z}}akowicz et al.}{2013}]{mol13} Molenda{-\.{Z}}akowicz J. et al., 2013, MNRAS, 434, 1422
\bibitem[\protect\citeauthoryear{Neilsen et al.}{2015}]{neil15} Neilsen M. B., Schunker H., Gizon L., Ball W. H., 2015, A\&A, 582, A10
\bibitem[\protect\citeauthoryear{Paxton et al.}{2011}]{bb13} Paxton, B., Bildsten, L., Dotter, A., Herwig, F., Lesaffre, P., Timmes, F., 2011, ApJS, 192, 3
\bibitem[\protect\citeauthoryear{Paxton et al.}{2013}]{Paxton et al.} Paxton B. et al., 2013, ApJS, 208, 49
\bibitem[\protect\citeauthoryear{Perryman}{2015}]{Perrymanp15} Perryman M. A. C. Lindegren L., Kovalevsky J., et al. 1997, A\&A, 323, 49
\bibitem[\protect\citeauthoryear{Pinsonneault et al.}{2014}]{pin14} Pinsonneault M. H. et al., 2014, ApJS, 215, 19
\bibitem[\protect\citeauthoryear{Porto de Mello\& da Silva}{1997}]{porsil97} Porto de Mello G. F. \& da Silva L., 1997, ApJ, 482, 289
\bibitem[\protect\citeauthoryear{Sahlholdt et al.}{2018}]{Sahlholdt} Sahlholdt, C., Aguirre, V., Casagrande, L. et al., 2018, MNRAS, 476, 1931
\bibitem[\protect\citeauthoryear{Sharma et al.}{2016}]{Shr16} Sharma S., Stello D., Bland-Hawthorm J., Huber D., Bedding T. R., 2016, ApJ, 822, 15
\bibitem[\protect\citeauthoryear{Stello et al.}{2008}]{Stello} Stello D., Bruntt H., Preston H., Buzasi D., 2008, ApJ, 674, L53
\bibitem[\protect\citeauthoryear{Skumanich}{1972}]{Sku72} Skumanich A., 1972, ApJ, 171, 565
\bibitem[\protect\citeauthoryear{Thoul}{1972}]{Thoul} Thoul A. A., Bahcall J. N., Loeb  A., 1994, ApJ, 421, 828
\bibitem[\protect\citeauthoryear{van Saders et al.}{2016}]{vansadersall} van Saders J. L., Ceillier T., Metcalfe T. S., Silva Aguirre V., Pinsonneault M. H., Garc{\'{i}}a R. A., Mathur S., Davies G. R., 2016, Nature, 529, 181
\bibitem[\protect\citeauthoryear{White et al.}{2017}]{white16} White T. R. et al., 2017, A\&A, 601, A82 
\bibitem[\protect\citeauthoryear{yildiz2011}{2011}]{paper2011} Y{\i}ld{\i}z M., 2011, MNRAS, 412, 2571 
\bibitem[\protect\citeauthoryear{PaperI}{2014a}]{paperI} Y{\i}ld{\i}z M., \c{C}elik Orhan Z., Aksoy  C., Ok S., 2014a, MNRAS, 441, 2148 (Paper I)
\bibitem[\protect\citeauthoryear{yildiz2014b}{2014b}]{yildiz2014} Y{\i}ld{\i}z M., \c{C}elik Orhan Z., Kayhan  C., Turkoglu G. E., 2014b, MNRAS, 445, 4395
\bibitem[\protect\citeauthoryear{PaperII}{2015}]{PaperII} Y{\i}ld{\i}z M., \c{C}elik Orhan Z., Kayhan C., 2015, MNRAS, 448, 3689 (Paper II)
\bibitem[\protect\citeauthoryear{PaperIII}{2016}]{PaperIII} Y{\i}ld{\i}z M., \c{C}elik Orhan Z., Kayhan C., 2016, MNRAS, 462, 1577 (Paper III)
\bibitem[\protect\citeauthoryear{PaperIII}{2017}]{PaperIII} Y{\i}ld{\i}z M., \c{C}elik Orhan Z., \"Ortel S., Roth M., 2017, MNRAS, 470, 25
\bibitem[\protect\citeauthoryear{Zinn et all}{2018}]{Zinn} Zinn J. C., Pinsonneault M. H., Huber D., Stello D., 2018, ApJ, 878, 2
\end{thebibliography}
\end{document}